\documentclass[fleqn,usenatbib]{mnras}

\usepackage{newtxtext,newtxmath}
\usepackage[T1]{fontenc}

\DeclareRobustCommand{\VAN}[3]{#2}
\let\VANthebibliography\thebibliography
\def\thebibliography{\DeclareRobustCommand{\VAN}[3]{##3}\VANthebibliography}

\usepackage{graphicx}
\usepackage{amsmath}
\usepackage{natbib}
\usepackage[utf8]{inputenc}
\usepackage{CJKutf8}

\title[H\,\textsc{i} in COLIBRE galaxies]{Atomic hydrogen in $z=0$ galaxies in the COLIBRE simulations}

\author[K.~A.~Oman et al.]{Kyle A. Oman$^{1,2}$\thanks{E-mail: kyle.a.oman@durham.ac.uk},
  Ruby J. Wright$^{3}$,
  Andrea Gebek$^{4,5}$,
  Diego Dado$^{1,2}$,
  Katherine E. Harborne$^{1,2}$,
  \newauthor
  Evgenii Chaikin$^{1,6}$,
  Carlos S. Frenk$^{1}$,
  Filip Hu\v{s}ko$^{6}$,
  Claudia D. P. Lagos$^{3}$,
  Aaron D. Ludlow$^{3}$,
  \newauthor
  Sylvia Ploeckinger$^{7}$,
  Alexander J. Richings$^{8,9}$,
  Matthieu Schaller$^{6}$,
  Joop Schaye$^{6}$
  and James W. Trayford$^{10}$
  \\
  $^{1}$Institute for Computational Cosmology, Physics Department, Durham University, South Road, Durham DH1 3LE, United Kingdom\\
  $^{2}$Centre for Extragalactic Astronomy, Physics Department, Durham University, South Road, Durham DH1 3LE, United Kingdom\\
  $^{3}$International Centre for Radio Astronomy Research, University of Western Australia, 7 Fairway, Crawley, WA 6009, Australia\\
  $^{4}$Department of Physics and Astronomy, Universiteit Gent, Proeftuinstraat 86 N3, B-9000 Ghent, Belgium\\
  $^{5}$Department of Physics, University of Bath, Claverton Down, Bath, BA2 7AY, United Kingdom\\
  $^{6}$Leiden Observatory, Leiden University, PO Box 9513, 2300 RA Leiden, the Netherlands\\
  $^{7}$Department of Astrophysics, University of Vienna, T\"{u}rkenschanzstrasse 17, A-1180 Vienna, Austria\\
  $^{8}$Centre for Data Science, Artificial Intelligence and Modelling, University of Hull, Cottingham Road, Hull, HU6 7RX, United Kingdom\\
  $^{9}$E. A. Milne Centre for Astrophysics, University of Hull, Cottingham Road, Hull, HU6 7RX, United Kingdom\\
  $^{10}$School of Physics, Engineering \& Computer Science, University of Hertfordshire, College Lane, Hatfield AL10 9UW, United Kingdom
}

\date{Accepted XXX. Received YYY; in original form ZZZ}
\pubyear{\the\year{}}

\begin{document}
\label{firstpage}
\pagerange{\pageref{firstpage}--\pageref{lastpage}}
\maketitle

\begin{abstract}
    We present an overview of the properties of atomic hydrogen (H\,\textsc{i}) in galaxies at redshift $z=0$ in the COLIBRE cosmological galaxy formation simulations. COLIBRE directly simulates the multi-phase interstellar medium (ISM) and includes new methods to follow radiative cooling to $\approx 10\,\mathrm{K}$, non-equilibrium chemistry coupled to dust, and dust formation, growth \& destruction.
    We focus on the $(200\,\mathrm{cMpc})^3$ volume with resolution elements of $2\times 10^{6}\,\mathrm{M}_{\odot}$.
    No H\,\textsc{i} properties informed the calibration of model parameters and yet we find very good agreement with the observed $M_{\star}$-$M_{\mathrm{HI}}$ relation, the H\,\textsc{i} mass and velocity width functions, the H\,\textsc{i} mass-size relation, and (baryonic) Tully-Fisher relations, albeit with some mismatches at the extremes of the galaxy population.
    H\,\textsc{i} surface density profiles are realistic, except at large radii around massive ($M_{\star}\gtrsim 10^{10}\,\mathrm{M}_{\odot}$) galaxies where we find a population of possibly spurious circumgalactic H\,\textsc{i} clouds (although their abundance and masses decrease with increasing mass resolution).
    Using smaller simulations to vary some of the uncertain model parameters, we find that the main scaling relations are at most mildly affected, although this may not extend to detailed H\,\textsc{i} morphokinematics.
   Convergence across mass resolution levels a factor of $8$ higher and lower is good: the median H\,\textsc{i} mass and H\,\textsc{i} `occupation fraction' as a function of halo mass, and the H\,\textsc{i} mass function, are remarkably stable, although the highest resolution model overproduces H\,\textsc{i} sources by a factor of $\approx2$ at $M_{\mathrm{HI}}\gtrsim 3\times 10^{9}\,\mathrm{M}_{\odot}$.
    Upturns in the low-mass and low-velocity slopes of the H\,\textsc{i} mass and velocity width functions with potentially subtle observational signatures are predicted.
\end{abstract}

\begin{keywords}
galaxies: general -- galaxies: ISM -- radio lines: ISM
\end{keywords}

\section{Introduction} \label{sec:intro}

Atomic hydrogen (H\,\textsc{i}) gas is a fundamental ingredient for galaxy formation. It is abundant in star-forming galaxies and fuels star formation through cooling into the molecular clouds that are the sites of ordinary star formation in galactic discs \citep{1971ApJ...163..155H,2006ApJ...650..933B,2009ApJ...693..216K,2011ApJ...728...88G,2017ApJ...848L...7C}. Simple time scale arguments show that the atomic phase is a dynamic one. Without constant replenishment of the atomic phase to feed the star-forming engine of a galaxy, it would run out of fuel: $M_\mathrm{HI}/\mathrm{SFR} < t_\mathrm{H}$ \citep[e.g.][]{2008A&ARv..15..189S,2011MNRAS.415...61S,2017ApJ...845..133S,2026MNRAS.548ag810T,2026MNRAS.548ag660L}, where $M_{\mathrm{HI}}$ is the H\,\textsc{i} mass, $\mathrm{SFR}$ is the star formation rate, and $t_{\mathrm{H}}$ is the Hubble time. Furthermore, star formation often drives strong, galaxy-scale winds that carry H\,\textsc{i} and metals out of the disc to mix with circumgalactic gas, likely with loading factors $\eta=\dot{M}_\mathrm{HI}/\mathrm{SFR}$ of order unity or more, implying even higher rates of feeding into the H\,\textsc{i} phase \citep[e.g.][]{2014MNRAS.442L.105M,2016MNRAS.462.3854L}. Depending on circumstances, outflowing gas can be heated and escape or induce cooling of surrounding gas \citep[e.g.][]{2022ApJ...924...82F}, potentially feeding back onto the cold disc in a galactic fountain \citep{2006MNRAS.366..449F,2008A&ARv..15..189S}.

H\,\textsc{i} is also notoriously sensitive to galaxy interactions. H\,\textsc{i} discs often extend well beyond the bright optical disc and are relatively weakly gravitationally bound, allowing them to respond easily and often dramatically to any perturbation (\citealp{2008A&ARv..15..189S} discuss a collection of examples). H\,\textsc{i} also provides the readily observable 21-cm spectral line that can serve as a kinematic tracer enabling spatially unresolved \citep[e.g.][]{1962AJ.....67..431R,2000ApJ...533L..99M,2018MNRAS.474.4366P,2019MNRAS.484.3267L,2026MNRAS.549ag574Y} or finely resolved dynamical mass estimates out to large radii \citep[e.g.][]{1981AJ.....86.1791B,1989A&A...223...47B,2001ARA&A..39..137S,2008AJ....136.2648D,2016AJ....152..157L,2026IAUS..392..119H}.

The numerous and complex physical processes governing the formation, destruction/depletion (into ions or molecules) and transport of H\,\textsc{i} within and around galaxies make modelling it challenging, especially where it is to be explicitly spatially resolved as in the context of hydrodynamical simulations of galaxy formation and evolution. The hot ionized, warm neutral, cold neutral and molecular gas phases are often intermingled within galaxies. If a simulation does not explicitly resolve the relevant physical processes and scales, then a `subgrid' approach to modelling the multiphase interstellar medium (ISM) is required. Constrained by computational resource considerations dictating the achievable resolution and detail achievable for cooling, chemistry and other processes, all hydrodynamical galaxy formation simulations that follow representative populations of galaxies to $z=0$ must currently resort to such subgrid models \citep[see e.g. the review in][]{2026arXiv260625359L}. Notable exceptions that do explicitly aim to directly simulate the multi-phase ISM at the expense of the total volume simulated include NewHorizon \citep{2021A&A...651A.109D} and FIREBox \citep{2023MNRAS.522.3831F}. The COLIBRE simulations studied in this work are the first to directly simulate the multiphase ISM in large volumes to $z=0$.

Simulations not aiming to directly simulate the multiphase ISM have adopted various approaches to approximating its thermodynamics and structure. Two-phase models \citep[e.g.][the model of the former is used in e.g. Illustris and IllustrisTNG; \citealp{2014MNRAS.444.1518V,2018MNRAS.473.4077P,2022A&A...661A..17B}]{2003MNRAS.339..289S,2010MNRAS.405.1491M} assume that each resolution element carries a fractional contribution from a hot and cold phase in pressure equilibrium. This enables the phases to coexist spatially but also forces them to share kinematics at the particle level. For H\,\textsc{i} this leads to the `cold ISM' inheriting many properties of hotter gas. Similar issues arise in the single-fluid approach of \citet[][]{2008MNRAS.383.1210S} where gas cooling into the cold phase is kept artificially hot and pressurised according to an effective equation of state. Production implementations \citep[e.g. EAGLE;][see also MUFASA and SIMBA; \citealp{2016MNRAS.462.3265D}, for a similar approach]{2015MNRAS.446..521S} limit cooling to minimum temperatures of the order of $10^{4}\,\mathrm{K}$. While this is appropriate for H\,\textsc{i} in the warm ISM, the cold phase again behaves as hotter gas. The approaches usually necessitate post-processing prescriptions to estimate the H\,\textsc{i} fractions of particles \citep[e.g.][]{2012MNRAS.420.2799D,2015MNRAS.452.3815L,2018ApJS..238...33D}. Empirical calibration and/or uncertain assumptions used to construct these prescriptions lead to ambiguity in the abundance and distribution of H\,\textsc{i} in the simulations.

Keeping the ISM artificially hot as in the approaches above is motivated by avoiding the artificial fragmentation of cooling gas when the Jeans' scale is unresolved \citep{1997ApJ...489L.179T}. However, \citet{2024MNRAS.528.2930P} have shown that since the gravitational force driving fragmentation is typically softened on scales larger than the Jeans' length of the cold ISM, and the thermal Jeans' length is usually smaller than the turbulent Jeans' scale because turbulence in the cold phase is usually supersonic, numerical fragmentation can be avoided through suitable choices of force softening and hydrodynamic smoothing scales.

A key issue once gas is allowed to cool into the cold phase is modelling the transition between the atomic and molecular phases, which in turn depends on the treatment of dust that acts as the main formation site for molecules. The chemistry of the ISM is sufficiently complex that modelling all species and reactions `on the fly' in large-volume simulations is computationally prohibitive. Assuming chemical and ionization equilibrium can greatly alleviate this, but non-equilibrium effects are important. Non-equilibrium treatments were pioneered by \citet{2009ApJ...697...55G} and \citet{2012MNRAS.425.3058C}; current solvers include Grackle \citep{2017MNRAS.466.2217S}, CHIMES \citep{2014MNRAS.440.3349R,2014MNRAS.442.2780R}, KROME \citep{2014MNRAS.439.2386G} and RAMSES-RTZ \citep{2022MNRAS.512..348K}. Recently, \citet{2025MNRAS.543..891P} presented the HYBRID-CHIMES model that combines a non-equilibrium treatment of hydrogen, helium and their contribution to free electrons with an equilibrium treatment of metal ions, while \citet{2026MNRAS.545f2040T} presented a dust model designed to track its formation and destruction `on the fly' and that is fully coupled to gas-phase chemistry. The reduced computational cost compared to larger chemical networks makes larger simulations feasible. \citet{2026MNRAS.547ag425P} cautions that treating oxygen species in equilibrium affects reactions involving H$_2$ and pushes the H\,\textsc{i} to H$_{2}$ transition to lower density by about an order of magnitude for temperatures of $\gtrsim 300\,\mathrm{K}$. This choice saves nearly a factor of $2$ in computational cost for a simulation as a whole, and we will show that it does not affect the H\,\textsc{i} mass function or H\,\textsc{i} mass-halo mass relation. 

These and other recent advances in ISM modelling have been applied in the recent COLIBRE simulations \citep{2026MNRAS.548ag375S}. These represent the current frontier for simulations resolving a multi-phase ISM in cosmologically representative volumes. We describe their key features in this context in Sec.~\ref{sec:sims}. We then undertake a first dedicated investigation of the H\,\textsc{i} properties of the galaxy populations in the simulations, focusing on $z=0$ with an emphasis on comparisons with observational 21-cm data. A complementary study of the cosmic abundance of H\,\textsc{i} and its evolution across cosmic history in COLIBRE simulations is in preparation by A.~J.~Richings et al., and a study of the thermodynamic structure of the ISM and interplay between the warm and cold neutral media in COLIBRE simulations can be found in \citet{2026arXiv260911654H}. \citet{2026MNRAS.549ag947L} showed that the COLIBRE simulations accurately reproduce the $z=0$ H\,\textsc{i} and H$_{2}$ Kennicutt-Schmidt relations between gas and star formation rate surface densities, including their scatter and secondary correlations with stellar surface density, and make predictions for and compare with available observations at higher redshift. \citet{2026MNRAS.550g1319L} showed that COLIBRE reproduces observed size-mass relations for a range of definitions at $z=0$ and out to at least $z=1.5$, as well as angular momentum-mass relations and `baryonic' mass-size and angular momentum-mass relations \citep{2025A&A...703A.223H} that incorporate H\,\textsc{i} gas.

This article is structured as follows. We give our first visual and qualitative impressions in Sec.~\ref{sec:visual-impressions}. We then investigate the scaling of H\,\textsc{i} mass with halo and stellar mass (Sec.~\ref{sec:scaling}), the H\,\textsc{i} mass and velocity width functions (Sec.~\ref{sec:himf-hiwf}), the H\,\textsc{i} content of massive, quiescent galaxies (Sec.~\ref{sec:massive-quenched}), the H\,\textsc{i} mass-size relation (Sec.~\ref{sec:mass-size}), the Tully-Fisher and baryonic Tully-Fisher relations (Sec.~\ref{sec:tfr-btfr}), and H\,\textsc{i} surface density profiles (Sec.~\ref{sec:SigmaHI}). We follow up on two features identified in Sec.~\ref{sec:visual-impressions} in Secs.~\ref{sec:central-holes} \& \ref{sec:cgm-clouds}. We summarize our findings in Sec.~\ref{sec:conclusions}.

\section{The COLIBRE simulations} \label{sec:sims}

Full details of the COLIBRE simulations are given in \citet{2026MNRAS.548ag375S} and references therein, with details of the model calibration given in \citet{2026MNRAS.548ag300C}. Here we summarize key features relevant to H\,\textsc{i} in galaxies.

The simulations were carried out using the \textsc{swift} code \citep{2024MNRAS.530.2378S} and the SPHENIX smoothed particle hydrodynamics scheme \citep{2022MNRAS.511.2367B}. Second-order Lagrangian perturbation theory initial conditions were generated at $z=63$ with the \textsc{monofonic} code \citep{2021MNRAS.503..426H}. The adopted $\Lambda$CDM cosmological parameters are from the Dark Energy Survey Year 3 \citep[DES Y3;][]{2022PhRvD.105b3520A} results, specifically the `3$\times$ 2pt $+$ All Ext.' constraints. Initial conditions include $4$ dark matter particles for each baryonic particle, giving the tracers of the two species approximately equal mass. The increased number of dark matter particles (relative to the commonly-used equal number of dark matter and gas particles in the initial conditions) suppresses spurious energy transfer \citep{2021MNRAS.508.5114L,2023MNRAS.525.5614L} between the two that artificially increases galaxy stellar sizes and velocity dispersions.

Gas radiative cooling and heating are modelled using HYBRID-CHIMES \citep{2025MNRAS.543..891P}, based on the chemical network solver CHIMES \citep{2014MNRAS.440.3349R,2014MNRAS.442.2780R}, and include non-equilibrium effects for $9$ hydrogen and helium species, and free electrons associated with their ions. Other species ($147$, for a total of $157$ species and $11$ elements tracked in all) are treated in chemical equilibrium, except that the metal cooling rates are corrected for the non-equilibrium free electron density. Each element and its ions contribute independently to metal-line cooling. The simulations feature a fully coupled, live dust model \citep{2026MNRAS.545f2040T} that tracks the formation and destruction of $3$ chemical species and $2$ grain sizes. Dust is coupled to the chemical network and can catalyse reactions, notably the formation of H$_{2}$.

Radiative transfer is not modeled explicitly, but some aspects are approximated. Cooling radiation propagates freely (except for CO, H$_2$O and OH, where self-absorption is included in HYBRID-CHIMES), but photoionization, photodissociation and photoheating rates include local radiative transfer approximations. The Jeans length\footnote{The maximum of the thermal and turbulent Jeans lengths, where the turbulent case assumes a 1D turbulent velocity dispersion of $6\,\mathrm{km}\,\mathrm{s}^{-1}$.} times the volume densities of H\,\textsc{i}, H$_2$, He\,\textsc{i}, He\,\textsc{ii}, CO and dust approximate their respective column densities \citep{2001ApJ...562L..95S}, which are used to compute attenuation corrections to rates. The Jeans length approximation is appropriate for self-gravitating gas but not pressure-confined clouds. At the resolutions used in COLIBRE (see below), pressure-confined clouds are often marginally or not resolved, while warm self-gravitating discs are resolved \citep{2026MNRAS.548ag375S}, so this approximation is at least justifiable \textit{a priori}.

The Jeans length approximation approach to estimating surface densities is also used to define the local interstellar radiation field (ISRF) and cosmic ray density. The ISRF intensity and cosmic ray density are assumed to be proportional to the local star formation rate surface density, which is in turn assumed to be proportional to the local gas surface density with a power law index of $1.4$ as observed by \citet{1998ApJ...498..541K}. The spectral shape of the ISRF is fixed to that given in \citet{1987ASSL..134..731B}. The normalisation of the ISRF is set to that from the `weak ISRF' model from \citet{2025MNRAS.543..891P}. We refer the interested reader to that work and \citet[][sec~3.1]{2026MNRAS.548ag375S} for further details.

H\,\textsc{ii} regions ionized by young stars are usually not resolved in COLIBRE simulations. They are modelled by stochastically labelling a subset of gas particles in the immediate vicinity of young stellar particles as H\,\textsc{ii} regions. These are assumed to contain singly-ionized hydrogen and helium, and are prevented from cooling below $10^{4}\,\mathrm{K}$ and from forming stars. Details of the H\,\textsc{ii} region model are given by \citet{2026MNRAS.546ag268B}.

The metagalactic UV \& X-ray background radiation sourced by galaxies and quasars is modelled using a modified version of the model from \citet{2020MNRAS.493.1614F}. Refer to \citet{2020MNRAS.497.4857P}, \citet{2025MNRAS.543..891P} and \citet[][sec.~3.1]{2026MNRAS.548ag375S} for further details.

COLIBRE's combination of gas cooling to low temperatures in large-volume simulations and explicit prediction of H\,\textsc{i} fractions of particles self-consistently coupled to the rest of the chemical network and dust model provides a currently unique dataset where H\,\textsc{i} can be studied in representative samples of galaxies without resorting to additional prescriptions to identify or `label' H\,\textsc{i} gas; the distribution and properties of H\,\textsc{i} gas are fully specified as predictions of the model. Nevertheless, significant uncertainties remain, particularly around the treatment of radiation shielding and the interstellar radiation field. These translate into corresponding uncertainties around the abundance and distribution of H\,\textsc{i} in the simulations. The COLIBRE suite includes additional simulations that vary parameters including the interstellar radiation field strength, the shielding length for ionizing radiation, and others (these will be described in detail by Chaikin~et~al., in preparation). We will use these to assess the sensitivity of our findings to these model uncertainties (Appendix~\ref{app:vars}) -- in brief, most model predictions that we investigate in this work are remarkably insensitive to reasonable variations of key model parameters.

Star formation occurs in gas that satisfies a local gravitational instability criterion depending on density and local thermal and turbulent velocity dispersion \citep{2024MNRAS.532.3299N}. Stellar feedback includes pre-supernova modes from stellar winds, radiation pressure and H\,\textsc{ii} regions \citep{2026MNRAS.546ag268B}, and kinetic and thermal energy injection from supernovae \citep{2012MNRAS.426..140D,2023MNRAS.523.3709C}. Black holes grow \citep{2022MNRAS.516..167B,2022MNRAS.515.4838N}, and feed back energy according to two model variants. The fiducial COLIBRE model used for the largest volume runs uses a purely thermal active galactic nuclear (AGN) feedback model \citep{2009MNRAS.398...53B}. Smaller volumes are also simulated with both this model and a `hybrid' black hole spin-dependent thermal and kinetic jet model \citep{2026MNRAS.547ag324H}. In this work we focus on the fiducial `Thermal-AGN' model.

Simulations were carried out at $3$ resolution levels with mean initial gas resolution element masses $m_{\mathrm{g}}$ of $1.47 \times 10^{7}$, $1.84 \times 10^{6}$ and $2.30 \times 10^{5}\,\mathrm{M}_{\odot}$, labelled `m7', `m6' and `m5'. The corresponding mean dark matter particle masses are $1.94\times10^{7}$, $2.42\times10^{6}$ and $3.03\times10^{5}\,\mathrm{M}_{\odot}$, and the gravitational softening lengths are $1.4$, $0.7$ and $0.35\,\mathrm{kpc}$ at $z=0$. The largest simulations completed at the time of writing\footnote{Simulations of $(50\,\mathrm{cMpc})^{3}$ and $(100\,\mathrm{cMpc})^{3}$ volumes at the highest resolution level are in progress at the time of writing.} at each resolution level have comoving volumes\footnote{We denote comoving megaparsecs `$\mathrm{cMpc}$'.} of $(400\,\mathrm{cMpc})^{3}$, $(200\,\mathrm{cMpc})^{3}$ and $(25\,\mathrm{cMpc})^{3}$. We use these for most of our analysis, but also make some use of smaller volume runs to check for box-size effects, and when specialist data products (e.g. radiative transfer post-processing) not yet available for the largest runs are needed. We focus most of our analysis on the intermediate (m6) resolution simulations. Simulations are labelled according to their box size and resolution level following \citet{2026MNRAS.548ag375S}, for instance L200m6 -- which we use as a fiducial reference model for most of our analysis -- is the $(200\,\mathrm{cMpc})^{3}$, m6-resolution run implementing the fiducial `Thermal-AGN' model.

For the m7 resolution level, the stellar and AGN feedback were calibrated \citep{2026MNRAS.548ag300C} to reproduce the $z=0$ galaxy stellar mass function \citep{2022MNRAS.513..439D} and size-mass relation \citep{2022MNRAS.509.3751H} over the interval $10^{9}<M_{\star}/\mathrm{M}_{\odot}<10^{11.3}$ using Gaussian process emulators trained on about $200$ smaller volume simulations. Independent of the emulator-based calibration, the AGN feedback coupling efficiency was adjusted to match the $z\approx0$ observationally-inferred supermassive black hole masses in massive galaxies. For the higher m6 and m5 resolution levels parameters were adjusted manually to achieve excellent convergence for the calibration targets across resolution levels.

Galaxies in the simulations were identified and catalogued in a three-stage process. First, friends-of-friends groups of dark matter particles linked by connections of at most $0.2$ times the mean initial interparticle spacing are identified \citep{1985ApJ...292..371D}. Non-dark matter particles are assigned to the group of their nearest dark matter particle, provided that it is within a halo and also within the linking length, at this stage. Each group is then processed using the \textsc{hbt-herons} substructure finder \citep{2012MNRAS.427.2437H,2018MNRAS.474..604H,2025MNRAS.543.1339F} that is informed by subhalo membership at earlier times and includes a gravitational binding criterion. All galaxies are labelled `central' at the first time when they are identified; when multiple central galaxies later become grouped within a single friends-of-friends group, the most massive is labelled `central' and the others `satellites' if the next most massive is at most $20$~per~cent less massive. Otherwise, the galaxy with the lowest orbital kinetic energy in the friends-of-friends group rest frame is the `central'. Finally, all galaxies are processed with the Spherical Overdensity Aperture Processor \citep[SOAP;][]{2025JOSS...10.8252M} that tabulates integral quantities using particles in a variety of spherical and projected apertures, including either all or only gravitationally bound particles.

Except where otherwise noted, galaxy stellar masses ($M_{\star}$), H\,\textsc{i} masses ($M_{\mathrm{HI}}$) and (instantaneous) star formation rates ($\mathrm{SFR}$) refer to the sum of that quantity over all particles in a $50\,\mathrm{kpc}$ spherical aperture identified by \textsc{hbt-herons} as gravitationally bound to that galaxy (i.e. excluding those identified as bound to satellite or companion galaxies). We have verified that our results are not strongly sensitive to this aperture choice, for example adopting a larger $100\,\mathrm{kpc}$ aperture would not lead us to different conclusions. Galaxy centres are defined as the location of their most-bound particle, and their rest frames are defined by the mean motion of all gravitationally-bound particles. Halo masses ($M_{\mathrm{200c}}$) are defined as the mass enclosed within a sphere within which the mean density is $200$ times the critical density $\rho_{\mathrm{crit}}=3H^{2}/8\pi G$, where $H$ is the Hubble-Lema\^{i}tre constant and $G$ is the gravitational constant. All analysis uses the final $z=0$ snapshots of the simulations.

\section{Atomic Hydrogen in COLIBRE} \label{sec:results}

We begin by forming an initial impression of H\,\textsc{i} in galaxies in the COLIBRE simulations guided by highly-resolved mock observations of a few example galaxies in Sec.~\ref{sec:visual-impressions}. We then present our main analysis in Sec.~\ref{sec:pop-props}, where we first follow up on features noted in the visual inspection of galaxies in Sec.~\ref{sec:visual-impressions} and then examine important H\,\textsc{i} properties of galaxies and scaling relations involving H\,\textsc{i}.

We note that the $z=0$ thermodynamic structure of the interstellar medium including its distribution as a function of density, temperature and pressure, and how H\,\textsc{i} populates this parameter space is discussed by \citet[][sec.~6.1, figs.~8 \& 9]{2026MNRAS.548ag375S}. Those authors also compare the $z=0$ H\,\textsc{i} gas fractions ($M_{\mathrm{HI}}/M_{\star}$) of COLIBRE galaxies as a function of stellar mass with observations (their sec.~7.2.3 and fig.~19). We discuss the same relation in Sec.~\ref{sec:scaling} and reach much the same conclusion that it is in very good agreement with observational constraints, but frame our discussion in the context of the rest of our analysis.

\subsection{Visual impressions} \label{sec:visual-impressions}

\begin{figure*}
  \includegraphics[width=\textwidth]{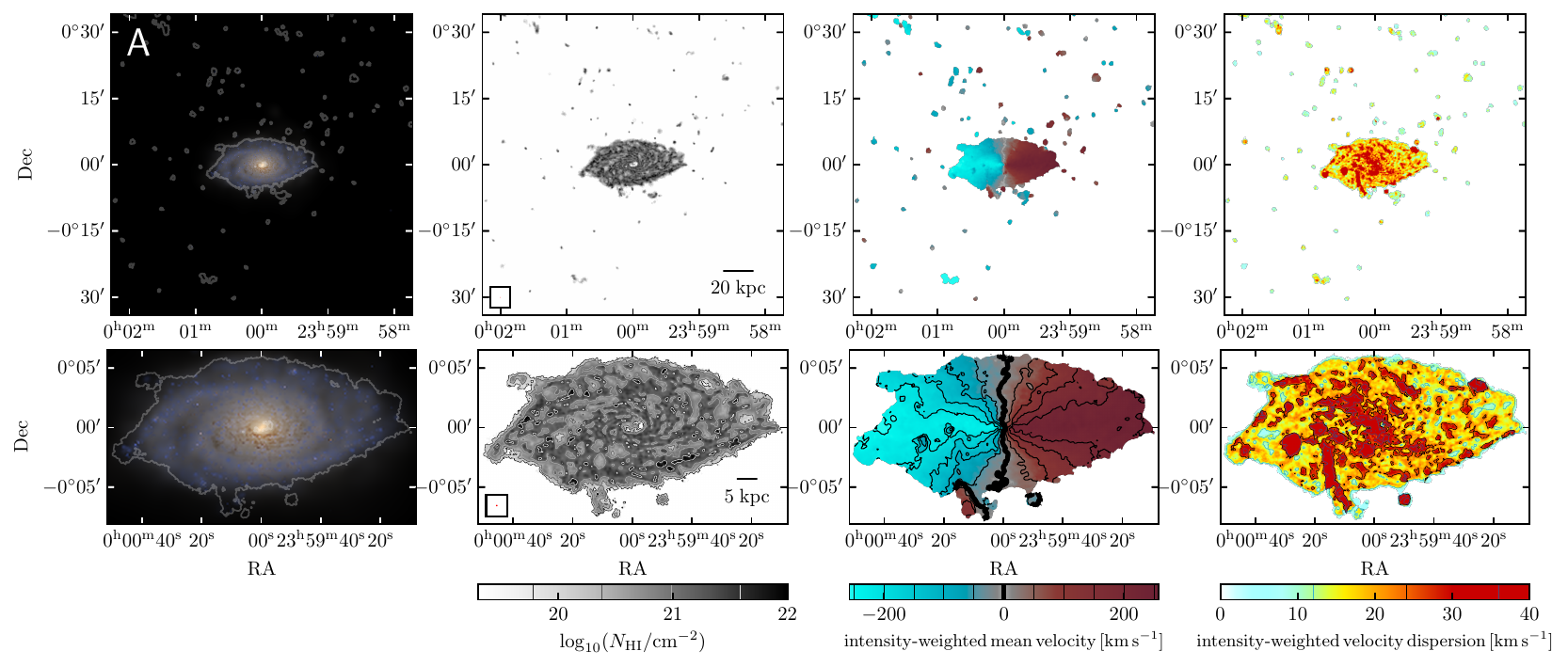}
  \includegraphics[width=\textwidth]{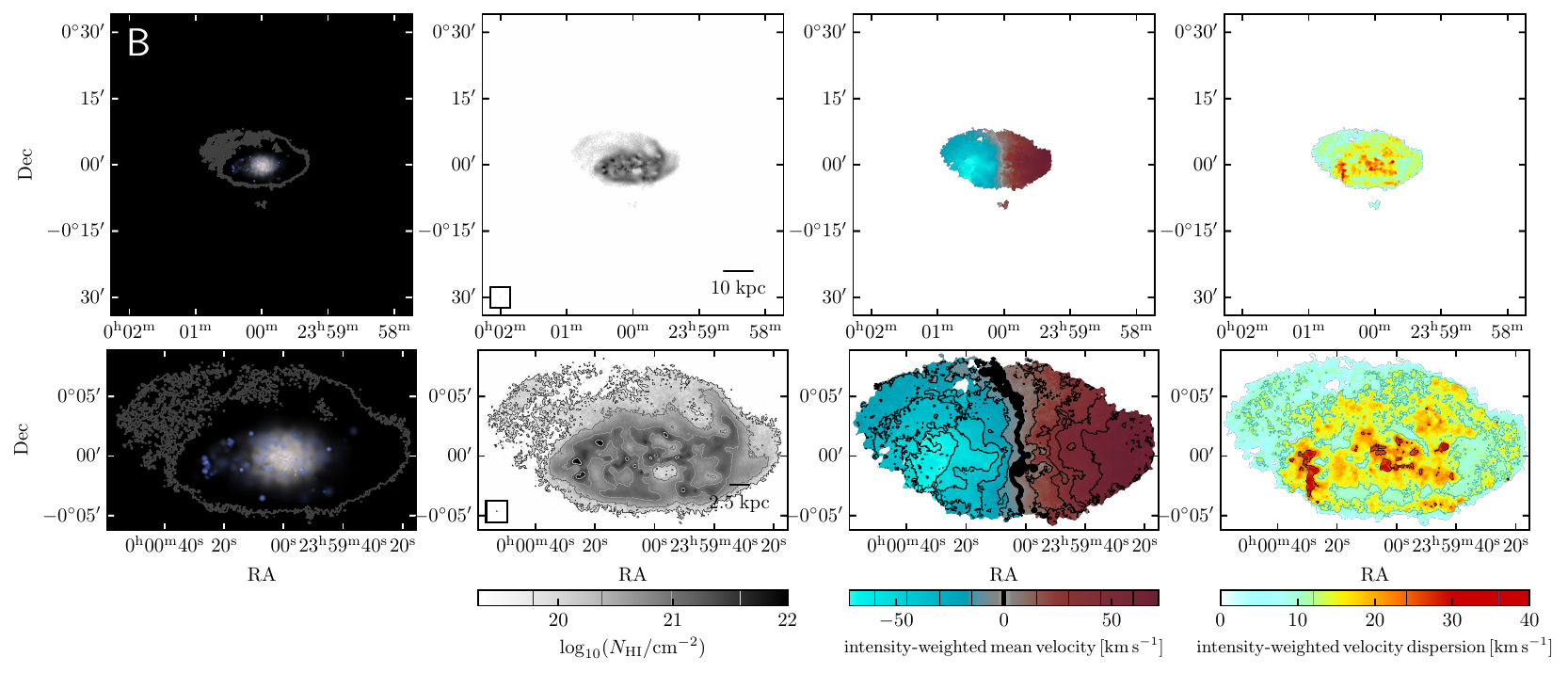}
  \caption{Examples of H\,\textsc{i} discs from the COLIBRE L025m5 run `observed' similarly to the highest-resolution (\texttt{r00\_t00}) reduction of MHONGOOSE survey data from MeerKAT: $8\,\mathrm{arcsec}$ and $1\,\mathrm{km}\,\mathrm{s}^{-1}$ resolution, $1\,\mathrm{deg}$ field of view and column density sensitivity of $10^{19.8}\,\mathrm{cm}^{-2}$ at $3\sigma$ over $16\,\mathrm{km}\,\mathrm{s}^{-1}$ -- further details are given in Appendix~\ref{app:mocks-mhongoose}. The upper panels (A) show a giant galaxy and the lower panels (B) a dwarf -- note the different spatial scales shown in the second column; further properties of the galaxies are given in Table~\ref{tab:img-props}. Both are seen $\approx 60\,\mathrm{deg}$ inclined. In each group of panels (A \& B) the upper row shows from left to right: mock HST image with limiting surface brightness of $1\,\mathrm{L}_{\odot}\,\mathrm{pc}^{-2}$ and H\,\textsc{i} column density contour at $10^{19.8}\,\mathrm{cm}^{-2}$; H\,\textsc{i} column density map; first moment map (intensity-weighted mean velocity) centred on the systemic velocity of the disc; second moment map (intensity-weighted velocity dispersion). The second row zooms in to show only the primary source, masked in 3D, and adds contours to the moment maps. The small red point in the box in the second panel shows the beam size. The outermost column density contour repeats the contour in the left panel; additional contours are spaced by factors of $4$. Velocity contours are spaced by $50$ and $15\,\mathrm{km}\,\mathrm{s}^{-1}$, respectively. Velocity dispersion contours are at $12$, $24$ and $36\,\mathrm{km}\,\mathrm{s}^{-1}$. The giant galaxy has a regularly rotating disc with a small central hole (see Sec.~\ref{sec:central-holes}) and dense clouds reaching $N_\mathrm{HI}\approx 10^{22}\,\mathrm{cm}^{-2}$. Much of the disc has a velocity dispersion of $\approx 15\,\mathrm{km}\,\mathrm{s}^{-1}$, but there are numerous local patches exceeding $30\,\mathrm{km}\,\mathrm{s}^{-1}$. The disc is surrounded by small clouds out to $>100\,\mathrm{kpc}$ broadly co-rotating with the disc (see Sec.~\ref{sec:cgm-clouds}). The dwarf's H\,\textsc{i} disc extends well beyond its stellar component, has a slightly lopsided and irregular morphology, but shows regular rotation and lower velocity dispersion, although there are still a few local peaks above $30\,\mathrm{km}\,\mathrm{s}^{-1}$. It lacks the population of small clouds that seem typical around more massive galaxies. The giant galaxy (A) can be interactively visualised at \url{colibre.strw.leidenuniv.nl/sliders.html}.}
  \label{fig:visual-impression-discs}
\end{figure*}

\begin{figure*}
  \includegraphics[width=\textwidth]{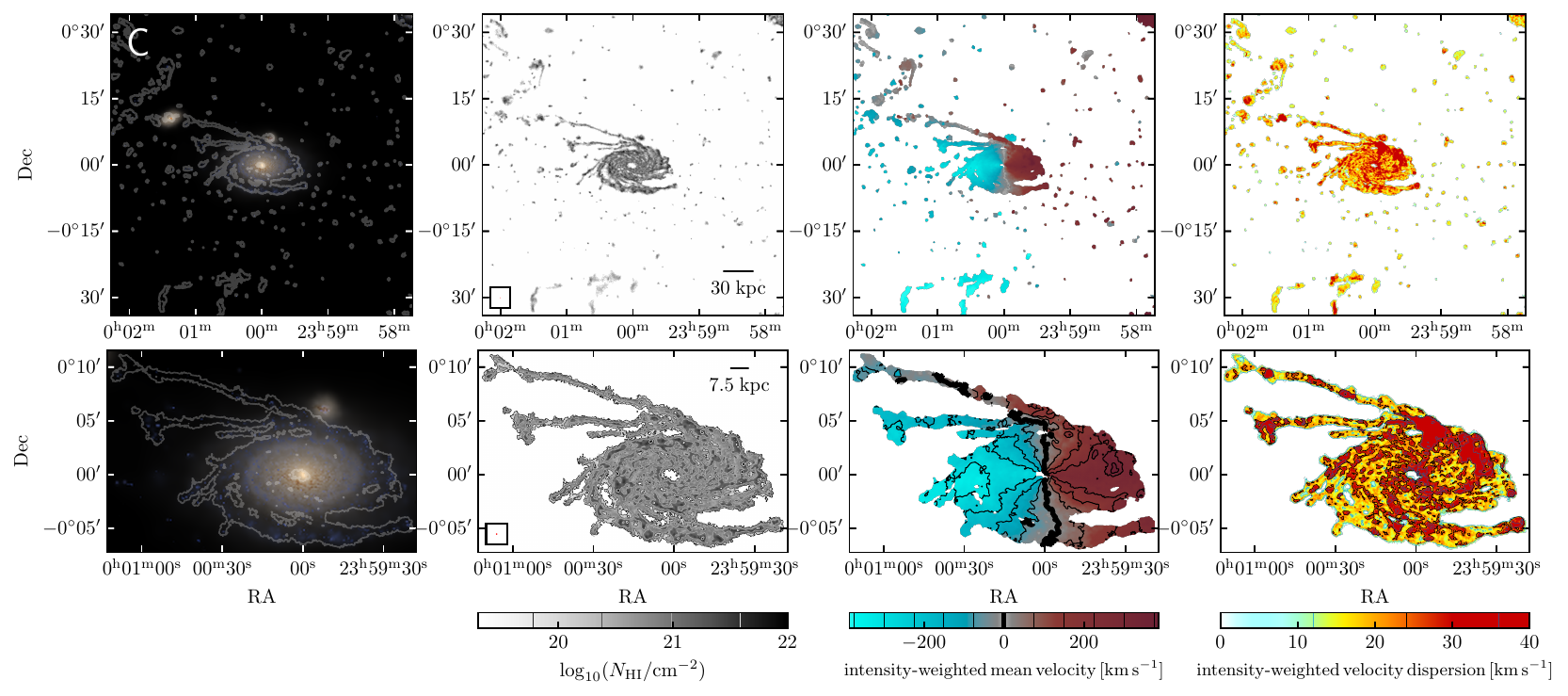}
  \includegraphics[width=\textwidth]{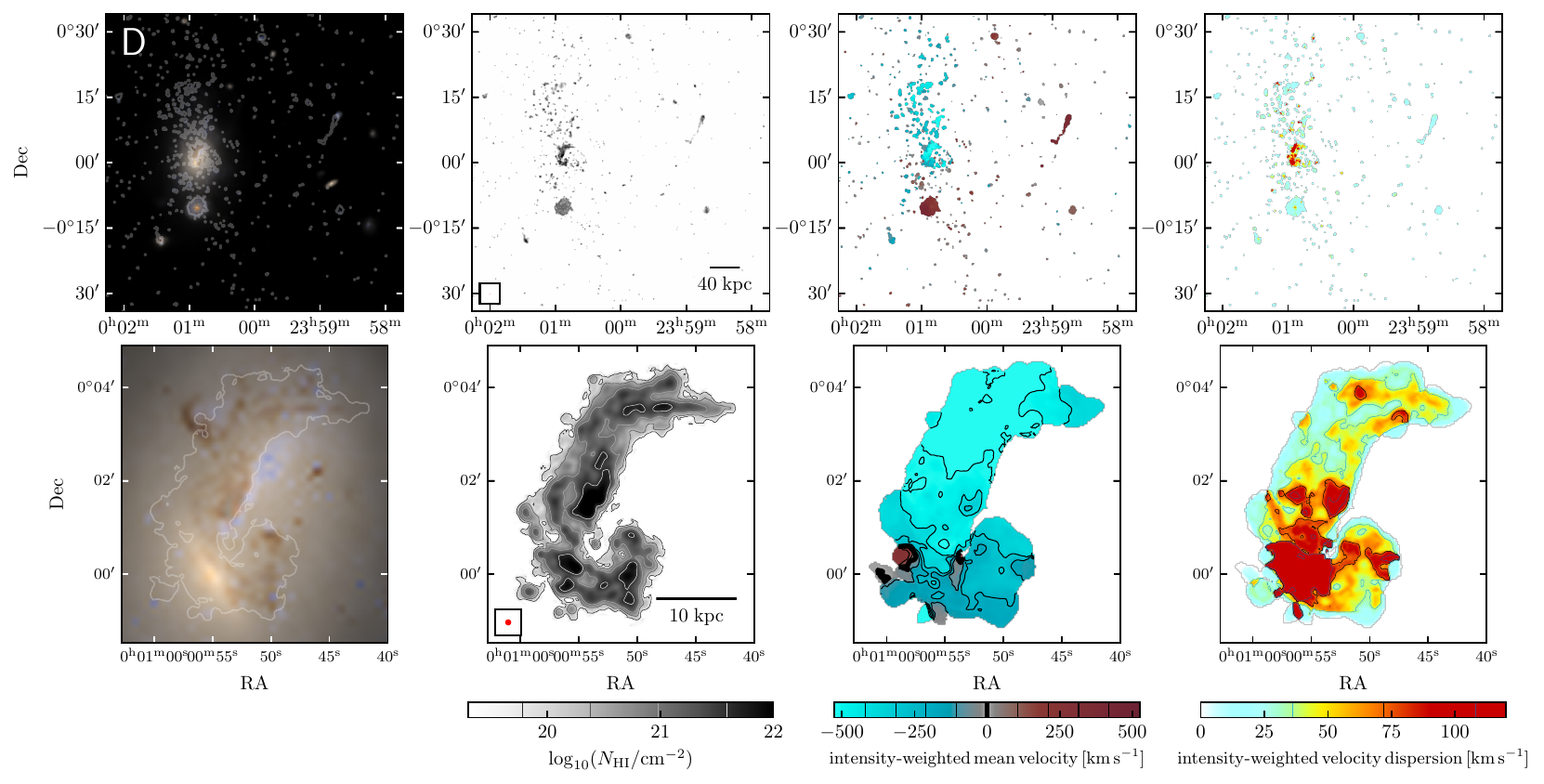}
  \caption{Similar to Fig.~\ref{fig:visual-impression-discs}, but showing two groups of interacting galaxies. The group in the upper panels (C) has a dominant central disc with a central hole, seen inclined $\approx 60\,\mathrm{deg}$. A prominent tail of gas connects it to a satellite to the north-east. Several other tails and clouds surround it, a few of which are associated with visible companions including one just north of the disc. The lower panels in the group again zoom in on the main source, masked in 3D (but not in the optical -- the companion visible there has associated H\,\textsc{i} gas visible in the upper panels, but this is masked out in the lower panels). The lower group of panels (D) shows a massive group (total mass $M_{\mathrm{200c}}\approx 2\times10^{13}\,\mathrm{M}_{\odot}$) dominated by an ongoing merger of two objects, and a disc, with many lower-mass companions. The image is centred on a point chosen to include most group galaxies in the field of view. The most massive galaxy (the merger shown in the zoom-in panels), highlighted in the lower panels, has very perturbed H\,\textsc{i} morphology and kinematics (the velocity dispersion contours and colour scale are increased by a factor of 3 for this system). Some of the satellites, such as one at the extreme north of the image and one with a long tail in the west-north-west, retain some H\,\textsc{i} gas, but many such, as the brightest group member in the south-west quadrant of the image, have none above the column density limit of the image. There are also many clouds without a bright ($M_\star\gtrsim10^{8}\,\mathrm{M}_{\odot}$) optical counterpart.}
  \label{fig:visual-impression-groups}
\end{figure*}

Before quantitatively assessing properties of the COLIBRE galaxy population in Sec.~\ref{sec:pop-props}, we first discuss our initial impression of the H\,\textsc{i} in COLIBRE galaxies drawn from the highest-resolution (L025m5) simulation. We adopt an observational point of view for this and inspire our approach from exceptionally deep and finely-resolved observations from the MHONGOOSE survey \citep{2024A&A...688A.109D}. To this end, we create mock-observed moment maps and complementary optical images of two disc galaxies (Fig.~\ref{fig:visual-impression-discs}) and two interacting systems: a disc with a massive satellite and a galaxy group (Fig.~\ref{fig:visual-impression-groups}). The galaxies are `observed' to a depth of $10^{19.8}\,\mathrm{cm}^{-2}$ ($3\sigma$ over $16\,\mathrm{km}\,\mathrm{s}^{-1}$) at $8\,\mathrm{arcsec}$ spatial and $1\,\mathrm{km}\,\mathrm{s}^{-1}$ velocity resolution. Targets are placed at distances of $5$-$20\,\mathrm{Mpc}$ (see Table~\ref{tab:img-props}, which also summarizes the basic properties of the most massive galaxy in each field) with a $1\,\mathrm{deg}$ field of view ($\approx 90$-$350\,\mathrm{kpc}$). This broadly mimics the \texttt{r00\_t00} reduction of the MHONGOOSE data, which is that with the highest spatial resolution which comes at the expense of some column density sensitivity. Optical images mimic HST colour composites with a limiting depth of $1\,\mathrm{L}_{\odot}\,\mathrm{pc}^{-2}$; these are more approximate than the H\,\textsc{i} mocks and intended only to give an overall impression of the appearance in the optical, e.g. the spatial resolution is not matched to HST. Full details of the mock observing techniques are given in Appendix~\ref{app:mocks-mhongoose}.

Field~A in Fig.~\ref{fig:visual-impression-discs} contains a massive ($M_\star\approx 10^{11}\,\mathrm{M}_{\odot}$, $M_\mathrm{HI}\approx 2\times 10^{10}\,\mathrm{M}_{\odot}$) spiral. The H\,\textsc{i} also forms (weak) spiral arms in a regularly rotating, cold (H\,\textsc{i} intensity-weighted line of sight velocity dispersion $\lesssim 20\,\mathrm{km}\,\mathrm{s}^{-1}$) gas disc -- here seen inclined by $\approx 60\,\mathrm{deg}$. There are, however, localised regions with elevated H\,\textsc{i} velocity dispersions \citep[$\approx 40\,\mathrm{km}\,\mathrm{s}^{-1}$, i.e. higher than typically observed, see e.g.][]{2010MNRAS.401.2113E,2015AJ....150...47I,2016AJ....151...15M} throughout the disc. Most of these are associated to star-forming regions or clouds close enough in velocity to the disc to be merged with it by the \textsc{SoFiA-2} source finding algorithm. Most of the disc is at an H\,\textsc{i} column density of about $10^{21}\,\mathrm{cm}^{-2}$, with localised clouds reaching nearly $10^{22}\,\mathrm{cm}^{-2}$. The galaxy has a central hole in its H\,\textsc{i} disc with a diameter of $\approx 5\,\mathrm{kpc}$. Such central holes are common (but not ubiquitous) in discs of about $M_{\star}=10^{10}\,\mathrm{M}_{\odot}$ and above (see also field~C in Fig.~\ref{fig:visual-impression-groups}). Such holes are not only devoid of H\,\textsc{i} gas but of all gas (not shown, but see Sec.~\ref{sec:central-holes}), i.e. the same feature is apparent in maps of the total gas density. They are therefore not a consequence of central cooling and condensation into the molecular phase. Instead, they are driven predominantly by AGN feedback activity; e.g. at fixed stellar mass, their prominence is correlated with the central supermassive black hole mass (Thompson et al., in preparation), and we verified that such holes do not occur in a small simulation (using the same initial conditions as L025m6) with AGN feedback switched off. The incidence rate of these holes is explored in Sec.~\ref{sec:central-holes}.

A second striking feature in field~A is the population of dense H\,\textsc{i} clouds ($N_{\mathrm{HI}}\lesssim 10^{21}\,\mathrm{cm}^{-2}$) surrounding the disc out to large radii (the field of view reaches almost $100\,\mathrm{kpc}$). There are $90$ `visible' clouds identified by the \textsc{SoFIA-2} source finder, and $461$ in all (see Sec.~\ref{sec:cgm-clouds}). None of these have optical counterparts with $M_{\star}>10^{6}\,\mathrm{M}_{\odot}$. A handful are in dark matter subhaloes and have optical counterparts just above the mass of a stellar resolution element ($\approx 2\times 10^{5}\,\mathrm{M}_{\odot}$), but most have no associated stars or dark matter subhalo. A minority ($5$ clouds) have non-zero star formation rates, and some ($10$ clouds) contain small amounts of molecular hydrogen. The smallest have sizes similar to the gravitational softening scale ($\approx 900\,\mathrm{pc}$). Most are co-rotating with the disc and therefore have likely cooled out of the CGM rather than being launched from the disc. Such populations of clouds are common around massive galaxies in COLIBRE, whether early- or late-type, but become less abundant with decreasing mass and resolution, and are not usually found around dwarfs ($M_{\star} \lesssim 10^{9}\,\mathrm{M}_{\odot}$). The incidence rate of similar cloud populations is estimated in Sec.~\ref{sec:cgm-clouds}.

Field~B in Fig.~\ref{fig:visual-impression-discs} contains a dwarf ($M_\star \approx 10^{9}\,\mathrm{M}_{\odot}$, $M_{\mathrm{HI}}\approx2\times10^{9}\,\mathrm{M}_{\odot}$) disc galaxy. The H\,\textsc{i} disc extends more than twice as far as the bright optical disc and has a slightly lopsided, asymmetric morphology. There is a prominent hole off-centre to the south-west, and another at a larger radius to the north-west -- these have a different origin from the holes discussed above and are associated with stellar feedback, not AGN activity. The population of clouds surrounding the more massive disc in field~A is conspicuously absent, except for one low-density ($N_{\mathrm{HI}}<10^{20}\,\mathrm{cm}^{-2}$) cloud just to the south of the disc. This particular dwarf has a velocity field reflecting its very regular rotation (again seen inclined by about $60\,\mathrm{deg}$), although this is not a generic feature of similar-mass galaxies in COLIBRE \citep{2026MNRAS.547ag356D}, and has low H\,\textsc{i} velocity dispersion ($\lesssim 15\,\mathrm{km}\,\mathrm{s}^{-1}$) throughout most of the disc except for a few localised regions with associated star-formation activity.

Field~C in Fig.~\ref{fig:visual-impression-groups} is another massive disc ($M_{\star}\approx2\times10^{11}\,\mathrm{M}_{\odot}$, $M_{\mathrm{HI}}\approx2\times 10^{10}\,\mathrm{M}_{\odot}$) and shares some similarities with the disc in field~A: it has weak spiral structure, regular rotation, a central hole and a population of clouds in the CGM co-rotating with the disc. However, in contrast to the relatively isolated disc in field~A, this one has multiple lower-mass companions clearly visible in the optical image and shows clear signs of interaction, especially with the galaxy to the north-east that has pulled out a long tail of H\,\textsc{i} following a close fly-by. Pockets of young blue stars are just visible in the two tails to the north-west and the tail to the south-east. Most patches of elevated ($>30\,\mathrm{km}\,\mathrm{s}^{-1}$) velocity dispersion throughout the disc correspond to star-forming regions and are therefore most likely due to supernova feedback, but the ongoing interaction likely also contributes.

Field~D is a more complex group ($M_{\mathrm{200c}} \approx 2\times10^{13}\,\mathrm{M}_{\odot}$) with multiple massive ($M_\star\gtrsim 10^{10}\,\mathrm{M}_{\odot}$) galaxies and many lower-mass members. The brightest object highlighted in the lower panels is actually two nearly equal-mass gas-rich galaxies in an ongoing merger according to the \textsc{hbt-herons} halo finder. These have a very irregular and kinematically perturbed morphology, as expected for such a merger. Some of the lower-mass satellites in the group have associated H\,\textsc{i}, but many do not, presumably having lost it to winds or stripping processes and being unable to re-accrete more in the group environment. There is one satellite in the west-northwest with a clear one-sided tail of H\,\textsc{i} characteristic of ram-pressure stripping, with another possible ram-pressure stripping candidate in the south-east. The group also contains very many (the majority of those visible) small H\,\textsc{i} clouds without stellar counterparts; these also have no associated dark matter subhaloes.

\begin{table*}
\caption{Basic properties of the most massive galaxy in each of fields A-D shown in Figs.~\ref{fig:visual-impression-discs} \& \ref{fig:visual-impression-groups}. All systems are drawn from the $z=0$ snapshot (number $127$) of the L025m5 COLIBRE simulation. The \textsc{hbt-herons} `Track ID' is a unique identifier; the subsequent columns give the stellar masses, H\,\textsc{i} masses, H$_{2}$ masses, instantaneous star formation rates and halo masses of the galaxies, and the distance at which they are placed to be mock-observed, respectively.}\label{tab:img-props}
\begin{tabular}{cccccccc}
\hline
&Track ID&$M_{\star}/\mathrm{M}_{\odot}$&$M_{\mathrm{HI}}/\mathrm{M}_{\odot}$&$M_{\mathrm{H2}}/\mathrm{M}_{\odot}$&$\mathrm{SFR}/\mathrm{M}_{\odot}\,\mathrm{yr}^{-1}$&$M_{\mathrm{200c}}/\mathrm{M}_{\odot}$&$D\,[\mathrm{Mpc}]$\\
\hline
A & 4705  & $9.3\times 10^{10}$ & $1.9\times 10^{10}$ & $2.6\times 10^{9}$ & 2.5  & $4.1\times 10^{12}$ & 10\\
B & 59399 & $9.4\times 10^{8}$  & $2.1\times 10^{9}$  & $2.8\times 10^{7}$ & 0.08 & $7.1\times 10^{10}$ & 5\\
C & 918   & $1.8\times 10^{11}$ & $2.2\times 10^{10}$ & $1.7\times 10^{9}$ & 1.6  & $9.5\times 10^{12}$ & 15\\
D & 5136  & $2.4\times 10^{11}$ & $8.9\times 10^{9}$  & $2.3\times 10^{8}$ & 0.5 & $2.2\times 10^{13}$ & 20\\
\hline
\end{tabular}
\end{table*}

\subsection{\texorpdfstring{H\,\textsc{i}}{HI} properties of the COLIBRE galaxy population} \label{sec:pop-props}

We next turn our attention to our main focus that is characterising the H\,\textsc{i} properties of galaxies in COLIBRE. This includes scaling relations between H\,\textsc{i}, halo and stellar mass and gas fractions (Sec.~\ref{sec:scaling}), the H\,\textsc{i} mass and velocity width functions (Sec.~\ref{sec:himf-hiwf}), the H\,\textsc{i} content of massive, quiescent galaxies (Sec.~\ref{sec:massive-quenched}), the H\,\textsc{i} mass-size relation (Sec.~\ref{sec:mass-size}), the Tully-Fisher and baryonic Tully-Fisher relations (Sec.~\ref{sec:tfr-btfr}), as well as H\,\textsc{i} surface density profiles (Sec.~\ref{sec:SigmaHI}). 

We will also explore two features that arose in the initial visual exploration presented in Sec.~\ref{sec:visual-impressions}: the central holes in H\,\textsc{i} discs (Sec.~\ref{sec:central-holes}), and the population of H\,\textsc{i} clouds surrounding some galaxies (Sec.~\ref{sec:cgm-clouds}).

\subsubsection{H\,\textsc{i} mass and gas fraction trends with stellar and halo mass} \label{sec:scaling}

\begin{figure}
  \includegraphics[width=\columnwidth]{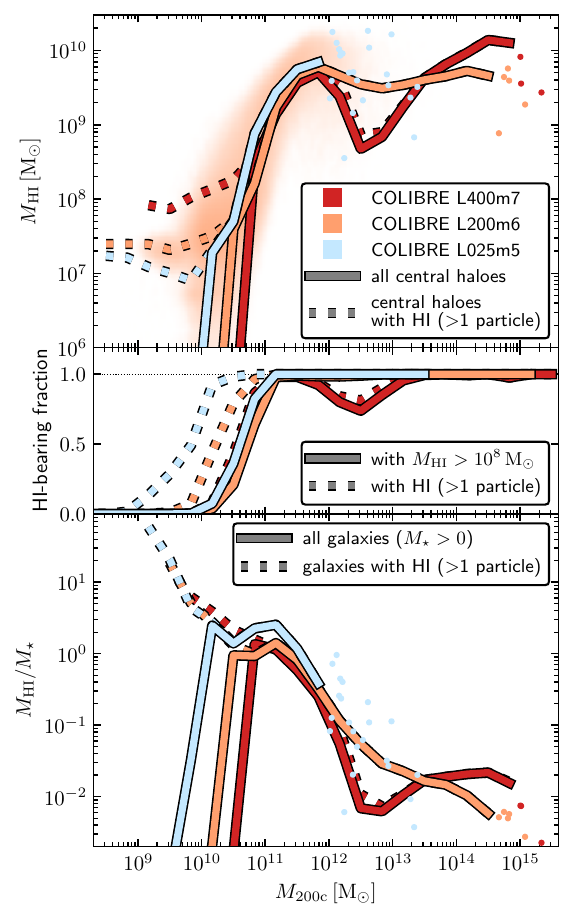}
  \caption{The H\,\textsc{i} mass-halo mass and related relations. \textit{Upper panel:} H\,\textsc{i} mass as a function of halo mass, $M_{\mathrm{200c}}$. The medians for all (central) haloes are shown with solid lines for each of the simulation resolution levels. Where there are fewer than $10$ haloes per $0.33\,\mathrm{dex}$ bin individual points are shown instead. The full distribution for L200m6 haloes is shown with the background shading. At low halo masses most haloes have less H\,\textsc{i} mass than one gas resolution element; the medians for haloes that have more than this are shown with dotted lines. \textit{Middle panel:} the fraction of haloes with at least $m_{\mathrm{g}}$ of H\,\textsc{i} is shown with the dotted lines to give context for the similar lines in the upper panel -- this is an explicitly resolution-dependent trend. The solid lines in this panel instead show the fraction of haloes with $M_{\mathrm{HI}}\geq 10^{8}\,\mathrm{M}_{\odot}$ -- a well-converged trend across all $3$ resolution levels. \textit{Lower panel:} H\,\textsc{i} gas fraction, $M_{\mathrm{HI}}/M_{\star}$, as a function of halo mass, $M_{\mathrm{200c}}$ for central galaxies (i.e. only central haloes with $M_{\star} > 0$). The solid lines show the median for all galaxies while dotted lines include only galaxies with at least one gas resolution element's mass in H\,\textsc{i}. There is a large population of low-mass galaxies without H\,\textsc{i}, but those with H\,\textsc{i} have high, nearly resolution-independent gas fractions.}
  \label{fig:Mhalo-MHI}
\end{figure}

We begin by examining the H\,\textsc{i} content of dark matter haloes as a function of their masses, shown in Fig.~\ref{fig:Mhalo-MHI}. The upper panel shows that this is reasonably well-converged with resolution for haloes with $10^{11}<M_{\mathrm{200c}}/\mathrm{M}_{\odot}<10^{12}$. At lower halo masses the median H\,\textsc{i} plummets as the majority of haloes do not contain even $m_{\mathrm{g}}$ of H\,\textsc{i}\footnote{Lower H\,\textsc{i} masses are possible because (i) a particle carries only about 70~per~cent hydrogen by mass, (ii) a particle may carry only a fraction of its hydrogen mass in H\,\textsc{i}, but in practice most particles have H\,\textsc{i} fractions close to either $0$ or to $1$, and (iii) gas particles may lose (or gain) some mass to black hole `nibbling' \citep{2022MNRAS.516..167B}.}, as illustrated in the second panel. Those low-mass haloes that do retain some H\,\textsc{i} keep around $10^{7}$-$10^{8}\,\mathrm{M}_{\odot}$ -- this is resolution-dependent, but by less than the factor of $8$ separating the masses of gas resolution elements. The fraction of haloes containing at least $10^{8}\,\mathrm{M}_{\odot}$ of H\,\textsc{i} is well-converged across all $3$ resolution levels, and the L200m6 and L025m5 simulations are similarly well-converged for haloes containing at least $10^{7}\,\mathrm{M}_{\odot}$ of H\,\textsc{i} (not shown). The large fraction of haloes with $M_{\mathrm{HI}}\approx0$ at $M_{\mathrm{200c}}\lesssim10^{11}\,\mathrm{M}_{\odot}$ and its dependence on resolution means that care must be taken when constructing galaxy samples that extend to low masses (in H\,\textsc{i}, halo, or stellar mass), taking into account the question being asked.

There is also a population of very H\,\textsc{i} gas-poor haloes at the high halo mass end, although these have enough H\,\textsc{i} that the fractions shown in the second panel of Fig.~\ref{fig:Mhalo-MHI} remain high. The dip seen here for the L400m7 simulation is attributed to a higher fraction of massive, quiescent galaxies in this simulation. The shape of the dip in the L400m7 simulation is consistent with the onset of effective AGN feedback above $M_{\mathrm{200c}}\approx 10^{12}\,\mathrm{M}_{\odot}$ modulating the H\,\textsc{i} content of galaxies \citep[see e.g.][fig.~5]{2020MNRAS.498...44C}. We have also verified that the location and depth of the dip correlates with the supermassive black hole mass at fixed halo mass. We will show below and later in Sec.~\ref{sec:himf-hiwf} that a sharper drop in the median H\,\textsc{i} mass at $M_{\mathrm{200c}}\gtrsim 10^{12.5}\,\mathrm{M}_{\odot}$ than obtained in the L200m6 and L025m5 simulations would likely be more consistent with observational constraints \citep[see also][sec.~7.2.3, and Sec.~\ref{sec:massive-quenched} in this work for further discussion of the H\,\textsc{i} content of these galaxies]{2026MNRAS.548ag375S}.

The lower panel of Fig.~\ref{fig:Mhalo-MHI} shows the H\,\textsc{i} gas fraction ($M_{\mathrm{HI}}/M_{\star}$) as a function of halo mass, necessarily including only central galaxies with $M_{\star}>0$. This mostly provides a variant view of trends discussed above, but we elaborate further on two points. First, the trend for L025m5 lies above those for the two lower-resolution simulations -- at most by a factor of $5$ at $M_{\mathrm{200c}}\approx 7\times10^{10}\,\mathrm{M}_{\odot}$, declining gradually to about $10$~per~cent at $M_{\mathrm{200c}}\approx10^{12}\,\mathrm{M}_{\odot}$, beyond which the L025m5 simulation samples too few haloes to make an accurate measurement. The galaxy stellar mass function is still rather well converged and in agreement with observations in this regime \citep[see][figs.~12 \& E1]{2026MNRAS.548ag375S}, providing a first hint that the L025m5 simulation is in some regimes overabundant in H\,\textsc{i} -- a theme that will recur and that we will explore more fully below. Possible physical reasons for this are also touched on by \citet{2026arXiv260911654H} in their sec.~4.1.3 and appendix~E.2; they find that the thermal pressure in the ISM does not converge \citep[see also][for a related discussion of the convergence of gas disc scale heights]{2026MNRAS.546ag268B}. Second, at the very low-mass end those few galaxies that do retain some H\,\textsc{i} gas (dotted lines) retain much more than $M_{\mathrm{HI}}=m_{\mathrm{g}}$, especially at higher resolution: where the lines terminate, most galaxies contain a single stellar resolution element, but several times (or several tens of times in the case of L025m5) more H\,\textsc{i} mass. Curiously the H\,\textsc{i} gas fraction ($M_{\mathrm{HI}}/M_{\star}$) of galaxies with $M_{\mathrm{HI}}\geq m_{\mathrm{g}}$ seems to converge rather well at the low-mass end across all resolution levels, even though the H\,\textsc{i} mass as a function of halo mass (upper panel) and H\,\textsc{i} gas fraction as a function of stellar mass (see below) do not.

The median H\,\textsc{i} content of haloes changes by less than $\approx 20$~per~cent when the intensity of the interstellar radiation field is increased by a factor of $10$ or switched off, when the shielding length for ionizing radiation is varied by a factor of $2$, or when additional non-equilibriun effects that are not modelled in the fiducial HYBRID-CHIMES configuration used for COLIBRE are modelled (see Appendix~\ref{app:vars}, Fig.~\ref{fig:Mhalo-MHI}).

\begin{figure}
  \includegraphics[width=\columnwidth]{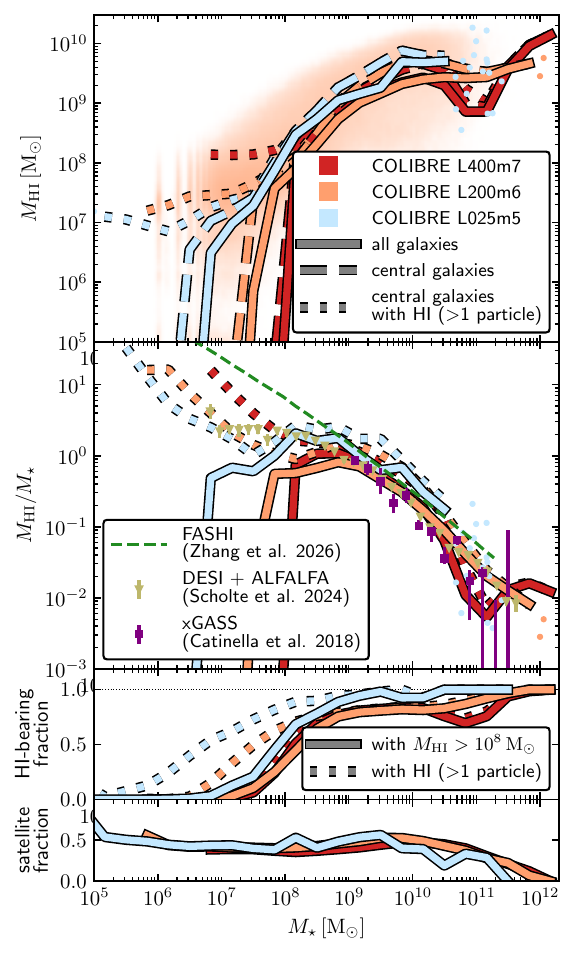}
  \caption{The H\,\textsc{i}-stellar mass and related relations. \textit{Upper panel:} H\,\textsc{i} mass as a function of stellar mass. The medians for all (central and satellite) galaxies are shown with the solid lines for each of the simulation resolution levels. Where there are fewer than $10$ haloes per $0.33\,\mathrm{dex}$ bin individual points are shown instead. The full distribution for L200m6 galaxies is shown with the background shading (opacity scales logarithmically). At low stellar masses most galaxies have less H\,\textsc{i} than the mass of one gas resolution element; the medians for galaxies with more H\,\textsc{i} than this are shown with dotted lines. Dashed lines show the median trends for only central (not satellite) galaxies. \textit{Second panel:} gas fraction ($M_{\mathrm{HI}}/M_{\star}$) as a function of stellar mass. Line styles are as in the upper panel (dashed lines omitted for clarity). Constraints on the median relation and uncertainty in the median from two stellar mass-selected surveys from \citet{2018MNRAS.476..875C} \& \citet{2024MNRAS.535.2341S} are shown with square and triangle markers with error bars. The median for an H\,\textsc{i}-selected, flux-limited survey from \citet{2026SCPMA..6929811Z} is shown with a thin dashed line. The COLIBRE simulations generally agree with these constraints, except at low masses where resolution effects become important (see text for a discussion). \textit{Lower panels:} Relations providing context for the upper panels. The third shows the fraction of galaxies with $M_{\mathrm{HI}}\geq m_{\mathrm{g}}$ (dotted lines) -- an explicitly resolution-dependent quantity. The fraction of galaxies with $M_{\mathrm{HI}}>10^{8}\,\mathrm{M}_{\odot}$ converges better, though not as well as the similar trend with halo mass (Fig.~\ref{fig:Mhalo-MHI}) -- this resolution dependence is therefore likely due to mildly unconverged stellar content across resolution levels. The bottom panel shows the fraction of satellite galaxies as a function of stellar mass.}
  \label{fig:Mstar-MHI}
\end{figure}

Fig.~\ref{fig:Mstar-MHI} is similar to Fig.~\ref{fig:Mhalo-MHI} but shows trends as a function of stellar mass and includes both central and satellite galaxies. The overall trends in $M_{\mathrm{HI}}$ as a function of $M_{\star}$ (upper panel) are similar to those as a function of $M_{\mathrm{200c}}$. The trends for all galaxies are reasonably well converged at intermediate masses ($10^{8}\lesssim M_{\star}/\mathrm{M}_{\odot}\lesssim 10^{10}$; although L025m5 is offset slightly to higher $M_{\mathrm{HI}}$), the median H\,\textsc{i} mass plummets at low mass reflecting bimodality in the H\,\textsc{i} content of galaxies, and a `dip' feature analogous to that at $M_{\mathrm{200c}}\approx10^{12.5}\,\mathrm{M}_{\odot}$ appears at $M_{\star}\approx10^{11}\,\mathrm{M}_{\odot}$ for the L400m7 simulation. When satellite galaxies are excluded, the median trends are offset to higher H\,\textsc{i} masses by up to a factor of about $3$ where the contribution of satellites is important (about half of all galaxies are satellites at $M_{\star}<10^{10.5}\,\mathrm{M}_{\odot}$, see the bottom panel of Fig.~\ref{fig:Mstar-MHI}).

The second panel of Fig.~\ref{fig:Mstar-MHI} shows the median gas fractions ($M_{\mathrm{HI}}/M_{\star}$) as a function of stellar mass, analogous to the lower panel of Fig.~\ref{fig:Mhalo-MHI}. Unlike the trend with halo mass where the gas fraction of low-mass galaxies with resolved gas fractions (dotted lines) converged very well across resolution levels, higher-resolution low-mass galaxies have systematically lower gas fractions. Since the trend is well-converged as a function of $M_{\mathrm{200c}}$ this reflects differences in the stellar-to-halo mass relation as a function of resolution, which is not surprising in this regime where the stellar content of galaxies is poorly sampled. The third panel of the figure reveals that the fraction of galaxies at fixed stellar mass that contain any H\,\textsc{i} is resolution dependent, but is again well-converged for galaxies with $M_{\mathrm{HI}}>10^{8}\,\mathrm{M}_{\odot}$, suggesting that selecting galaxies with H\,\textsc{i} masses corresponding to the mass of about $10$ resolution elements (i.e. about $10^{8}\,\mathrm{M}_{\odot}$ for the L400m7 simulation, and a factor of $8$ and $64$ less for the L200m6 and L025m5 simulations, respectively) is a lower bound below which resolution effects are clearly not negligible.

Observational constraints on the median gas fraction are also shown in the second panel of Fig.~\ref{fig:Mstar-MHI}. The measurements of \citet{2018MNRAS.476..875C} and \citet{2024MNRAS.535.2341S}, based on the xGASS\footnote{Extended GALEX (Galaxy Evolution Explorer) Arecibo SDSS (Sloan Digital Sky Survey) Survey} and DESI\footnote{Dark Energy Spectroscopic Instrument.} \& ALFALFA\footnote{Arecibo Legacy Fast ALFA (Arecibo L-band Feed Array).} surveys, respectively, are both stellar mass-selected samples and should therefore be compared with the solid lines in the figure. 

The agreement between COLIBRE and these observations is quantitatively excellent above $M_{\star}\approx 10^{9}\,\mathrm{M}_{\odot}$, especially for the L200m6 simulation. The L025m5 simulation is offset to gas fractions that are a factor of $2$-$4$ too high over most of the range $10^{8}<M_{\star}/\mathrm{M}_{\odot}<10^{11}$; near $M_{\star}\approx2\times10^{9}\,\mathrm{M}_{\odot}$ the agreement is better with an offset of about $50$~per~cent. This does not seem to be an artefact of the small simulation volume; the offset persists when comparing the L025m5 and L025m6 simulations instead. \citet{2024MNRAS.535.2341S} note a flattening of the slope below $M_{\star}\approx10^{9}\,\mathrm{M}_{\odot}$. COLIBRE qualitatively reproduces this, but the flattening is too pronounced in the L200m6 simulation. The slope change in COLIBRE reflects the onset of an increasing fraction of galaxies with $M_{\mathrm{HI}}\approx 0$ at low stellar masses (third panel of figure), so this may occur at stellar masses that are too high in the simulations. However, inspecting the dotted lines in the second panel (including only galaxies with $M_{\mathrm{HI}}\geq m_{\mathrm{g}}$) reveals that in the L200m6 simulation even these are lower than is observed -- the agreement is better for the L025m5 and L400m7 simulations. The gas fractions of mass gas-rich galaxies at $M_{\star}\lesssim10^{9}\,\mathrm{M}_{\odot}$ therefore seem slightly too low in the L200m6 simulation. The \citet{2024MNRAS.535.2341S} measurements are based on stacking of the H\,\textsc{i} signal so they may reflect the mean rather than the median, but we have checked that comparing to mean trends does not change this interpretation.

The median of an H\,\textsc{i}-selected sample from the FASH\,\textsc{i} survey \citep{2026SCPMA..6929811Z} is shown with the thin dashed line. This measurement is expected to lie above that for a stellar-mass selected sample because the H\,\textsc{i} selection biases the sample towards more gas-rich galaxies. The dotted lines for COLIBRE galaxies are also H\,\textsc{i} selected, but on a mass (not a flux) limit. Caution is therefore needed to make a comparison, but the $>1\,\mathrm{dex}$ offset around $M_{\star}\approx10^{7.5}\,\mathrm{M}_{\odot}$ (larger than the $\approx 0.5\,\mathrm{dex}$ $1\sigma$ scatter in the COLIBRE relation at this stellar mass scale -- not shown) may be difficult to reconcile despite the strong selection effects at play.

Overall, the gas fractions of COLIBRE galaxies are an excellent match to observations within the stellar mass range $10^{9}<M_{\star}/\mathrm{M}_{\odot}<10^{11.3}$ at all three resolution levels. We turn next to the global number densities of galaxies as a function of H\,\textsc{i} mass and spectral line width.

\subsubsection{The H\,\textsc{i} mass and velocity width functions} \label{sec:himf-hiwf}

The number densities of H\,\textsc{i}-detected galaxies as a function of their H\,\textsc{i} masses and velocity widths are among the most fundamental characterisations of the gas-rich galaxy population. Since the selection function of H\,\textsc{i} line surveys, such as ALFALFA \citep{2005AJ....130.2598G,2018ApJ...861...49H} that we use as a comparison benchmark in this section, is normally a joint function of $21\,\mathrm{cm}$ flux and line width\footnote{A narrower line compresses emission into fewer detector channels, leading to a higher signal-to-noise ratio at fixed total flux and enhancing detection probability.}, these two number density functions should be regarded as orthogonal marginalisations of the joint mass-line width function.

\begin{figure*}
  \includegraphics[width=\textwidth]{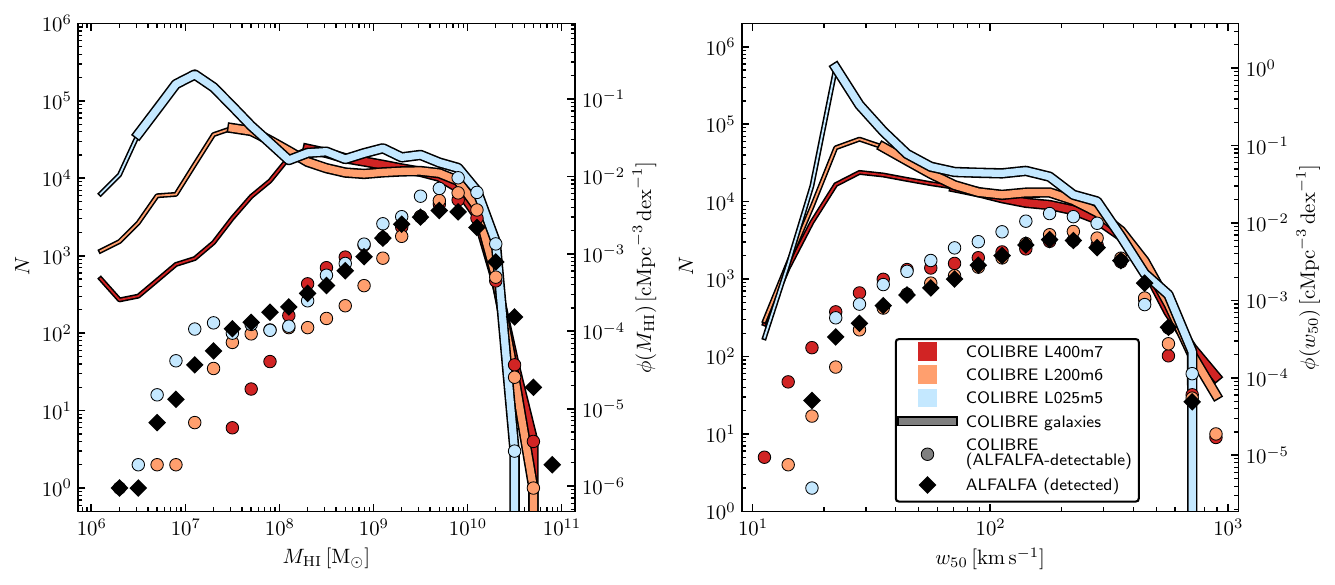}
  \caption{The COLIBRE H\,\textsc{i} mass function and velocity width function. \textit{Left panel:} the H\,\textsc{i} mass function of all galaxies ($M_{\star}>0$) in a conical mock survey matched in volume to the ALFALFA survey is shown for the $3$ simulation resolution levels with solid lines. The region with positive slope at low masses is strongly affected by the resolution, though the turnover of the peak near $10^{7}\,\mathrm{M}_{\odot}$ for L025m5 is likely resolved. The cutoff at high mass in L025m5 is a box size effect. The left axis marks galaxy counts within the volume; the right axis marks their number densities. The coloured points show galaxy counts in mock catalogues that closely mimic the ALFALFA selection function and can be directly compared to the black points drawn from the ALFALFA `$\alpha.100$' catalogue \citep{2018ApJ...861...49H}. The convergence across resolution levels and the agreement with observations are generally good but require careful interpretation (see Sec.~\ref{sec:himf-hiwf}). \textit{Right panel:} similar to the left panel but for the H\,\textsc{i} velocity width function. $w_{50}$ is measured as the full width at half maximum of forward-modelled H\,\textsc{i} spectra. The increasing height of the peak near $20\,\mathrm{km}\,\mathrm{s}^{-1}$ with increasing resolution, and to some extent the overall increasing normalization, is mostly a resolution effect. The cutoff at the high-$w_{50}$ end for L025m5 is a box size effect. The agreement with the shape of the observed number count distribution, and in the case of L200m6, also its normalisation, is remarkable but again requires careful interpretation (Sec.~\ref{sec:himf-hiwf}).}
  \label{fig:himf-hiwf}
\end{figure*}

In this section we compare the ALFALFA H\,\textsc{i} mass and velocity width functions to mock surveys of the L400m7, L200m6 and L025m5 COLIBRE simulations. A full description of the mock survey methodology is given in Appendix~\ref{app:mocks-alfalfa}. In brief, we select an arbitrary point in the simulation volume to place the `observer' and choose an arbitrary reference direction. We select a conical survey volume with its tip at the observer and a half-opening angle of $\approx46.8\,\mathrm{deg}$ around the reference direction such that the solid angle subtended is equal to ALFALFA's total solid angle of $6518\,\mathrm{deg}^{2}$, and extending to a maximum distance of $200\,\mathrm{Mpc}$. When necessary (mostly for the L025m5 simulation), we tile the periodic simulation volume to cover the survey volume\footnote{This has little influence on the results presented in this section except that for the L025m5 simulation: (i) bin-to-bin noise in the H\,\textsc{i} mass function is artificially amplified due to repeated galaxies with identical masses (the H\,\textsc{i} velocity width function is not affected because each galaxy has a different sight line and therefore a different-shaped spectrum as described below), and (ii) the high-mass and velocity width ends of the functions are artificially truncated because the small volume does not sample the long-wavelength modes of the matter power spectrum that lead to the formation of the largest overdensities.}. For each galaxy\footnote{The incidence of H\,\textsc{i} sources without optical counterparts is briefly discussed in Appendix~\ref{app:himf-hiwf}, and is the topic of a paper in preparation by F.~Turini et al.} with $M_{\star} > 0$ (both centrals and satellites) we select the particles flagged as gravitationally bound according to \textsc{hbt-herons} and record an H\,\textsc{i} mass and forward-modeled spectrum. The spectrum is evaluated along the sight line from the observer to the galaxy. We approximately account for source confusion by merging sources closely separated in angle and velocity (see Appendix~\ref{app:mocks-alfalfa} for details). This leads to a volume-limited catalogue. We then construct an approximately flux-limited catalogue by evaluating the ALFALFA detection probability for each galaxy using\footnote{There is a typographical error in their eq.~(A6): $2.248$ should be $2.448$.} eqs.~(A4)-(A6) from \citet{2022MNRAS.509.3268O}. When the probability is not $0$ or $1$ we compare with a random number drawn from a uniform distribution spanning the interval $[0, 1)$ to decide whether to include it in the catalogue. The catalogue is only approximately flux limited because the detection probability also depends on velocity width. We remove sources outside of the intervals $10^{6}<M_{\mathrm{HI}}/\mathrm{M}_{\odot}<10^{11}$ and $10<w_{50}/\mathrm{km}\,\mathrm{s}^{-1}<1000$, of which there are less than $30$ present in any mock survey volume, and less than $4$ that would otherwise be detectable. This matches the limits to which we plot the H\,\textsc{i} mass and velocity width functions and ensures that exactly the same galaxies contribute to both functions. The same cuts are applied to the ALFALFA catalogue.

We show the H\,\textsc{i} mass and velocity width functions in Fig.~\ref{fig:himf-hiwf}. We first consider the `intrinsic' functions based on the volume-limited catalogues, shown with the solid lines. The H\,\textsc{i} mass function shows broadly good convergence across the three resolution levels where galaxies have $M_{\mathrm{HI}}\gtrsim10m_{\mathrm{g}}$. The influence of the finite resolution of the simulations is evident by the approximately parallel trends with positive slope at low $M_{\mathrm{HI}}$ below about $2 \times 10^{8}$, $3 \times 10^{7}$ and $10^{7}\,\mathrm{M}_{\odot}$ for L400m7, L200m6 and L025m5, respectively. The cutoff is not sharp (e.g. it does not occur at the particle mass $m_{\mathrm{g}}$, which is 1-2 decades lower than each of these breaks) because the limit is a combination of the halo mass limit (at about $20$ times the dark matter particle mass) and the mass-dependent occupation fraction for H\,\textsc{i} in haloes shown in Fig.~\ref{fig:Mhalo-MHI}. We plot the functions with a thinner line where we assess them to be strongly affected by limited resolution. Above these resolution limits the different simulations agree reasonably well, although the L025m5 run is offset high by a factor of $\approx 2$ at $M_{\mathrm{HI}} > 2\times 10^{8}\,\mathrm{M}_{\odot}$. At the extreme high-mass end the differences are driven by box size effects as noted above. The H\,\textsc{i} mass function does not seem to depend sensitively on assumptions around the interstellar radiation field intensity, shielding length for ionizing radation, or unmodelled non-equilibrium effects in HYBRID-CHIMES (see Appendix~\ref{app:vars}, Fig.~\ref{fig:himf-mv}).

The upturn in the H\,\textsc{i} mass function below $\approx 10^{8}\,\mathrm{M}_{\odot}$ in the L025m5 (also hinted at in L200m6) seems to be a prediction of the COLIBRE model. It is reminiscent of a similar feature in the H\,\textsc{i} mass function in the EAGLE simulations \citep{2017MNRAS.464.4204C}, but that was caused by haloes accumulating H\,\textsc{i} and remaining just below the threshold for star formation (i.e. the galaxies were star-less or `dark') and was widely regarded as a consequence of choices in the modelling approach rather than a physical prediction. Fig.~\ref{fig:himf-hiwf} includes only galaxies with at least one star particle, so the same interpretation does not hold for the COLIBRE model. There are `dark' haloes around $M_{\mathrm{HI}}\approx 10^{7}\,\mathrm{M}_{\odot}$ (see Appendix~\ref{app:himf-hiwf}) but the majority of haloes in this regime contain stars. We will return to this feature below.

The H\,\textsc{i} velocity width function is well-converged, to a level similar to the H\,\textsc{i} mass function. At $w_{50} \gtrsim 50\,\mathrm{km}\,\mathrm{s}^{-1}$ the L400m7 and L200m6 models agree rather well, while the L025m5 model is again offset high by a factor of $2$-$3$ for $w_{50}<300\,\mathrm{km}\,\mathrm{s}^{-1}$. The `spike' near $w_{50}=20\,\mathrm{km}\,\mathrm{s}^{-1}$ is higher at higher resolution because it is predominantly made up of the low-$M_{\mathrm{HI}}$ galaxies that are not resolved at lower resolution. Because galaxies across a wide range in $M_{\mathrm{HI}}$ contribute to the H\,\textsc{i} velocity width function at fixed $w_{50}$, it is not expected to converge as resolution increases until all galaxies at a given fixed $w_{50}$ are resolved. We therefore regard the upturn towards low $w_{50}$ most prominently seen in the L025m5 model as a physical prediction of the COLIBRE model worth considering further, rather than as a symptom of poor numerical convergence.

We next turn our attention to the (approximately) flux-limited catalogues and a direct comparison with ALFALFA data. We make this comparison at the catalogue level because producing an observationally-selected catalogue from our volume-limited catalogue for the simulations requires only a single assumption: that the empirically determined survey selection function is accurately determined. The inverse problem of correcting the observed flux-limited catalogue for incompleteness is much more difficult and requires this assumption plus an estimator for galaxy clustering in the survey volume, and is sensitive to distance errors (especially for the H\,\textsc{i} velocity width function). In some regimes assumptions are also required for the shape of the intrinsic joint $M_{\mathrm{HI}}-w_{50}$ distribution \citep[see][around fig.~4, for a detailed discussion of these issues]{2022MNRAS.509.3268O}. In summary, comparing in the flux-limited catalogue space yields much more reliable conclusions.

Number counts from the ALFALFA catalogue \citep[][`Code 1' sources only]{2018ApJ...861...49H} are shown with black points in Fig.~\ref{fig:himf-hiwf}. The coloured points show the directly comparable counts from our COLIBRE mock surveys. For the H\,\textsc{i} mass function the agreement is very good outside of the simulation resolution-affected regime -- while there are differences of factors of up to $\approx 3$, the agreement is much closer than other cosmological hydrodynamical simulations have been able to achieve \citep[see e.g.][for the same figure for the EAGLE model; see also \citealp{2017ApJ...850...97B,2019MNRAS.482.5606D}, but these authors compare to completeness-corrected catalogues which carry additional uncertainties]{2024MNRAS.533...67O}. The H\,\textsc{i} velocity width function agrees exceptionally well in terms of its shape, although the L025m5 model is offset everywhere to higher number counts. This leads us to the conclusion that galaxies in the L025m5 model are overall somewhat too gas-rich which makes them more easy to detect in a (mock) survey, reinforcing our interpretation of Fig.~\ref{fig:Mstar-MHI}. Where it resolves the galaxy population well ($M_{\mathrm{HI}} \gtrsim 3 \times 10^{7}\,\mathrm{M}_{\odot}$, $w_{50} \gtrsim 40\,\mathrm{km}\,\mathrm{s}^{-1}$) the L200m6 simulation provides an excellent match to H\,\textsc{i}-selected galaxy samples, at least in terms of number counts. Furthermore, the relatively small discrepancies can be traced to a specific subset of the galaxy population, as we will see next.

\begin{figure*}
  \includegraphics[width=\textwidth]{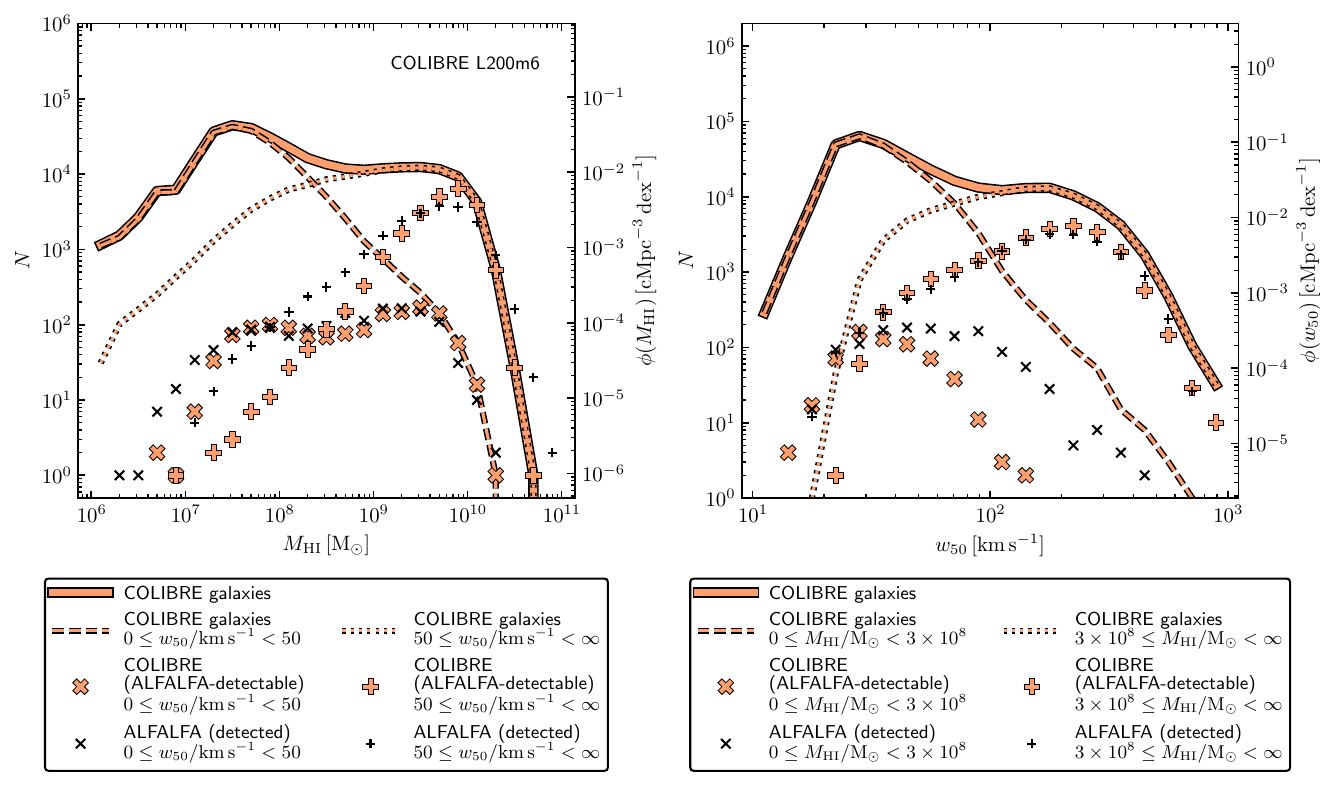}
  \caption{Similar to Fig.~\ref{fig:himf-hiwf}, the solid line for L200m6 is repeated from that figure. \textit{Left panel:} the H\,\textsc{i} is decomposed into galaxies with narrow ($w_{50}<50\,\mathrm{km}\,\mathrm{s}^{-1}$; dashed line and cross markers) and wide ($w_{50}>50\,\mathrm{km}\,\mathrm{s}^{-1}$; dotted line and plus markers) velocity widths and compared to the analogously decomposed observed number count distribution. \textit{Right panel:} as in the left panel, but splitting the H\,\textsc{i} velocity width function into low- and high-H\,\textsc{i} mass populations at $M_\mathrm{HI}=3 \times 10^{8}\,\mathrm{M}_{\odot}$. Taken together, these reveal that most of the tension with observations is driven by an underabundance of massive (i.e. large $w_{50}$), H\,\textsc{i}-poor galaxies in the simulations. This is compensated by an overproduction by a factor of nearly $2$ of high-$w_{50}$ galaxies with $M_{\mathrm{HI}} \approx 10^{10}\,\mathrm{M}_{\odot}$.}
  \label{fig:himf-hiwf-split-m6}
\end{figure*}

Fig.~\ref{fig:himf-hiwf-split-m6} again shows the H\,\textsc{i} mass and velocity width functions for the L200m6 simulation -- the solid lines are repeated from Fig.~\ref{fig:himf-hiwf}. We then bisect the H\,\textsc{i} mass function into two bins in velocity width (at $w_{50}=50\,\mathrm{km}\,\mathrm{s}^{-1}$), and the H\,\textsc{i} velocity width function into two bins of H\,\textsc{i} mass (at $3 \times 10^{8}\,\mathrm{M}_{\odot}$). We do this for both the volume-limited sample (shown with broken line styles) and the observationally-selected sample (shown with markers). The ALFALFA catalogue is also bisected in the same ways and shown with corresponding markers. This separates the volume-limited sample into two components around the inflections in the slopes of the two functions near $3 \times 10^{8}\,\mathrm{M}_{\odot}$ and $80\,\mathrm{km}\,\mathrm{s}^{-1}$. The comparison with the observed sample is particularly informative: the distribution of H\,\textsc{i} masses for low-velocity width galaxies (cross markers, left panel) is an almost perfect match to the observed catalogue, while there is a large deficit of $M_{\mathrm{HI}} < 10^{9}\,\mathrm{M}_{\odot}$ galaxies in the higher $w_{50}$ bin, and a smaller excess\footnote{Baker et al. (in preparation) find a population of early-type galaxies in COLIBRE that seem to be over-abundant in molecular gas, which may be related.} around $M_{\mathrm{HI}}=10^{10}\,\mathrm{M}_{\odot}$. It is therefore massive galaxies (large $w_{50}$) that are entirely responsible for the discrepancies seen in the left panel of Fig.~\ref{fig:himf-hiwf}. Similarly, for the H\,\textsc{i} velocity width function, there is excellent agreement with the data for galaxies in the higher H\,\textsc{i} mass bin, but a deficit of galaxies with broader lines for galaxies in the lower H\,\textsc{i} mass bin.

This leads us to conclude that the main shortcoming of the COLIBRE model in terms of faithfully reproducing number counts in H\,\textsc{i}-selected surveys is its deficit of H\,\textsc{i} gas-poor (but not H\,\textsc{i}-free) galaxies with broad lines and therefore large dynamical masses. Given the large dynamical masses, the galaxies must be resolved in the simulations, so the problem seems to be that they host too little H\,\textsc{i} gas.

We now return to the topic of the upturns at the low-$M_{\mathrm{HI}}$ and low-$w_{50}$ ends of the H\,\textsc{i} mass and velocity width functions. These are most obvious in the L025m5 simulation, but the slopes also get steeper at low $M_{\mathrm{HI}}$ and $w_{50}$ in the L200m6 simulation -- the L400m7 simulation does not adequately resolve the relevant scales. These upturns, if they were to occur in the observed galaxy population, would have significant consequences. In this scenario, about 70~per~cent of H\,\textsc{i}-bearing galaxies have $M_{\mathrm{HI}}<2 \times 10^{7}\,\mathrm{M}_{\odot}$, rather than only 30~per~cent if the shallower slope continues to low masses -- although such galaxies would still only account for 2~per~cent (rather than only 0.2~per~cent, based an integration of the H\,\textsc{i} mass function extrapolated to $M_{\mathrm{HI}}=10^{6}\,\mathrm{M}_{\odot}$) of the total H\,\textsc{i} mass in galaxies. The predicted upturn at low masses in the H\,\textsc{i} mass function has the clear observational signature of a shallower decline in number counts in the flux-limited sample towards lower masses: the trend is nearly flat around $10^{8}\,\mathrm{M}_{\odot}$ in the L200m6 and L025m5 simulations. Interestingly, a similar, though less pronounced, feature is present in about the same location in the ALFALFA catalogue. This suggests that a steepening slope of the (volume-limited) H\,\textsc{i} mass function at the low-mass end is a scenario that should be taken seriously and allowed for in future modelling efforts; the usual approach of single-Schechter function fitting \citep[e.g.][]{2005MNRAS.359L..30Z,2018MNRAS.477....2J,2025A&A...695A.241M} may be too simplistic. There are clear hints of a slightly steepening low-mass slope in the data shown in the cited references, but none comment on this possibility, likely because it occurs at a mass scale where uncertainties in the completeness corrections become large. The low-mass slope is also known to depend on local environment \citep{2005MNRAS.359L..30Z,2018MNRAS.477....2J,2020MNRAS.494.2090J}, which complicates the situation. \citet{2026SCPMA..6929811Z} recently published an even more accurate determination of the H\,\textsc{i} mass function using $>10^{5}$ FASH\,\textsc{i} detections and claim a constant slope at the low-mass end in spite of all points below $M_{\mathrm{HI}}=10^{7.5}\,\mathrm{M}_{\odot}$ that they include in their Schechter function fit lying significantly above the best-fitting line (see their fig.~12). Further investigation of the H\,\textsc{i} mass function guided by COLIBRE predictions will be worthwhile.

The sharp upturn in the low-$w_{50}$ end of the velocity width function predicted by COLIBRE leaves a less obvious signature in the flux-limited catalogue counts because many of the galaxies that contribute to it occupy a region of the $M_{\mathrm{HI}}-w_{50}$ plane that current surveys are blind to, particularly for the rise towards low line widths between $40$ and $80\,\mathrm{km}\,\mathrm{s}^{-1}$ \citep[see][fig.~4]{2022MNRAS.509.3268O}. There is a hint of an upturn below $w_{50}=45\,\mathrm{km}\,\mathrm{s}^{-1}$ in the ALFALFA data \citep[see fig.~1 in][]{2022MNRAS.509.3268O}. H\,\textsc{i}PASS did not reach these line widths and a velocity width function measurement based on FASH\,\textsc{i} data has not yet been published, although it has detected many more faint galaxies with narrow velocity widths than ALFALFA \citep[][fig.~4, see also \citealp{2026SCPMA..6929811Z}]{2025A&A...695A.241M} so new insights into this feature may be expected very soon.

\subsubsection{The H\,\textsc{i} content of massive, quiescent galaxies} \label{sec:massive-quenched}

Massive, quiescent galaxies are known to be H\,\textsc{i} and more broadly gas-poor objects, but measuring just how little H\,\textsc{i} they contain has been a long-standing challenge. Earlier surveys suffered either from shallow depth \citep[e.g.][]{2005AJ....130.2598G,2014MNRAS.443.1002L,2018ApJ...861...49H} or selection biases in e.g. morphology \citep[e.g.][]{2011MNRAS.416.1680C,2012MNRAS.422.1835S}. Recently, deep observations with FAST \citep{2026ApJ..1006..145L} have shown that a majority of massive ($10^{10}<M_{\star}/\mathrm{M}_{\odot}<10^{11.6}$), quiescent (colour $\mathrm{NUV}-r>5$ or specific star formation rate -- sSFR -- $\mathrm{SFR}/M_{\star}<10^{-11}\,\mathrm{yr}^{-1}$) galaxies have gas fractions $M_{\mathrm{HI}}/M_{\star}<10^{-2.4}$, with some upper limits reaching $10^{-3}$ and stacking implying an average for the non-detections below $10^{-3.5}$.

We showed in Sec.~\ref{sec:himf-hiwf} that the COLIBRE simulations seem to produce a deficit of massive (large $w_{50}$), gas-poor galaxies, but (i)~galaxies with no H\,\textsc{i} gas do not contribute to the mass and velocity width functions discussed above, and (ii)~that discussion was blind to ongoing or recent star formation activity. We therefore make a more direct comparison to the measurements of \citet{2026ApJ..1006..145L} here. Although their sample nominally includes galaxies in the stellar mass interval $10^{10}<M_{\star}/\mathrm{M}_{\odot}<10^{11.6}$, the low- and high-mass ends are very sparsely sampled, leading to a biased comparison when simulated galaxies are selected in the same range. We find that limiting the interval for both observations and simulations to $10^{10.2}<M_{\star}/\mathrm{M}_{\odot}<10^{11}$ results in very similar distributions of stellar masses and a fairer comparison, so we adopt this for our analysis. For the sSFR-selected sampled the stellar mass distributions are statistically indistinguishable (a Kolmogorov-Smirnov test yields $p\approx0.4$, unable to reject the null hypothesis that the samples are drawn from the same distribution), while for the colour-selected sample the simulations are biased ($p\approx10^{-3}$) to slightly lower stellar masses, by about $0.1\,\mathrm{dex}$. We include both central and satellite galaxies, as does the sample observed with FAST.

\begin{figure}
  \includegraphics[width=\columnwidth]{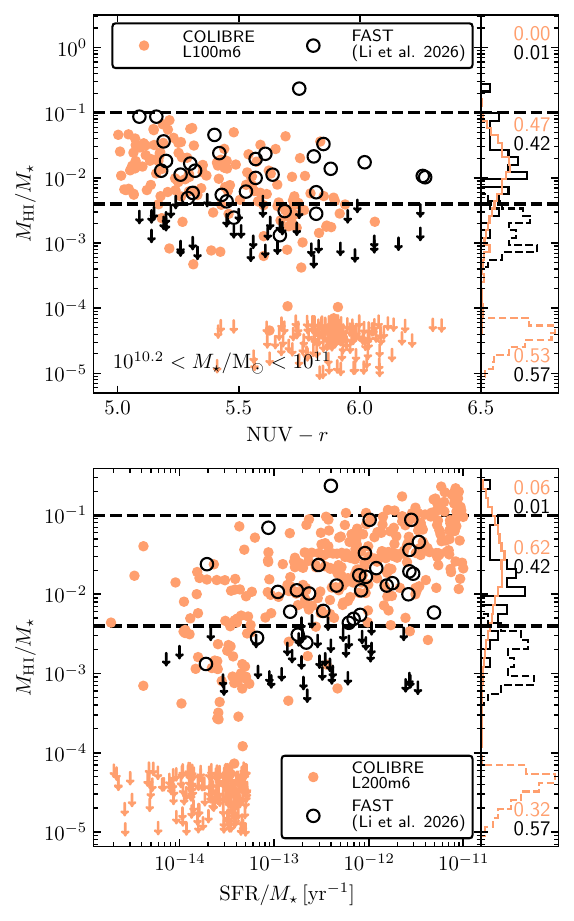}
  \caption{Distribution of H\,\textsc{i} gas fractions ($M_{\mathrm{HI}}/M_{\star}$) in massive, quiescent galaxies. All galaxies plotted have $10^{10.2}<M_{\star}/\mathrm{M}_{\odot}<10^{11}$. \emph{Upper panel:} H\,\textsc{i} gas fraction as a function of $\mathrm{NUV}-r$ colour. Galaxies are selected to be quiescent if $\mathrm{NUV}-r > 5$. Galaxies from the COLIBRE L100m6 simulation (filled points) are compared to recent deep FAST observations by \citet[][open points]{2026ApJ..1006..145L}. Observed upper limits are shown with downward arrows; COLIBRE galaxies with $M_{\mathrm{HI}}<m_{\mathrm{g}}$ are plotted as upper limits assuming that mass to evaluate their H\,\textsc{i} gas fraction. Broadband fluxes account for dust attenuation. The fractions of galaxies lying in the regions delineated by the dashed lines are labelled in the right sub-panel. Two thirds of the observed sample lie below the lower dashed line; 62~per~cent of the simulated galaxies fall in the same region. The gas fraction distributions of points and upper limits are plotted with solid and dashed lines in the right sub-panel and reveal very good agreement between the simulations and observations, except for a tail of high-$M_{\mathrm{HI}}/M_{\star}$, quiescent galaxies that is not observed. A random subsample ($1$ in $3$) of simulated galaxies are plotted for visual clarity, but all are used to construct the histograms. The combined (dashed and solid) histograms are normalised to unity. \emph{Lower panel:} Similar to the upper panel, but for galaxies selected to have specific star formation rates $\mathrm{SFR}/M_{\star}<10^{-11}\,\mathrm{yr}^{-1}$. Similar fractions of quiescent galaxies selected in this way have gas fractions $<4\times10^{-3}$ (i.e. $<10^{-2.4}$). We assign specific star formation rates of $(10^{-14}\,\mathrm{yr}^{-1}) \pm 0.7\,\mathrm{dex}$ to simulated galaxies with $<1$ particle's mass in H\,\textsc{i} to allow them to be shown on the figure. In this panel we use the L200m6 simulation and plot a random subsample of $1$ point in $24$.}
  \label{fig:massive-quenched}
\end{figure}

A faithful comparison with their measurements also requires broadband fluxes in the $r$ and $\mathrm{FUV}$ bands. These are tabulated in the \textsc{soap} catalogues that we use throughout our analysis. However, those values do not include any accounting for dust attenuation, which is particularly important in the $\mathrm{FUV}$. We therefore use dust-corrected $\mathrm{FUV}$ and $r$-band fluxes \citep{2026arXiv260714901G,2026MNRAS.tmp.1593L} computed with the radiative-transfer code \textsc{skirt} \citep{2020A&C....3100381C}. These are currently available for the L100m6 but not the L200m6 simulation. Therefore, unlike throughout most of this work, we make use of the smaller volume run when using these fluxes. When comparing in terms of star formation rates we use the L200m6 simulation.

We show the H\,\textsc{i} gas fraction ($M_{\mathrm{HI}}/M_{\star}$) in Fig.~\ref{fig:massive-quenched} for quiescent galaxies selected either with colour: $\mathrm{NUV}-r > 5$ (upper panel), or sSFR: $\mathrm{SFR}/M_{\star}<10^{-11}\,\mathrm{yr}^{-1}$ (lower panel). We note that these selections are not equivalent -- about twice as many galaxies ($1.6\times10^{-3}\,\mathrm{cMpc}^{-3}$) are included in the sSFR-selected sample as are in the colour-selected sample ($8.6\times10^{-4}\,\mathrm{cMpc}^{-3}$). Which is the fairer comparison is unclear -- on the one hand inferring physical quantities like stellar mass and star formation rate for observed galaxies is uncertain, but on the other hand the broadband fluxes are sensitive to the treatment of dust in the simulations, which also carries uncertainties.

Any correlation with either $\mathrm{NUV}-r$ colour or sSFR is weak, in agreement with the observations. We find a substantial population of massive, H\,\textsc{i}-poor quiescent galaxies in COLIBRE with gas fractions $<10^{-2.4}$ ($35$~per~cent of the quiescent population in the sSFR-selected sample, or $56$~per~cent in the colour-selected sample -- the latter in excellent agreement with the observed $57$~per~cent). Gas fractions reach as low as $10^{-5}$ with the caveat that one gas particle ($m_g\approx 10^{6}\,\mathrm{M}_{\odot}$) in a $M_{\star}=10^{10}\,\mathrm{M}_{\odot}$ galaxy implies a gas fraction of $10^{-4}$. We therefore treat H\,\textsc{i} masses of simulated galaxies of $10^{6}\,\mathrm{M}_{\odot}$ or less as upper limits. Nevertheless, the presence of a massive, H\,\textsc{i}-poor and quiescent galaxy population is an encouraging success of the model.

The average gas fraction of COLIBRE galaxies below the approximate sensitivity limit of most of the FAST observations ($M_{\mathrm{HI}}/M_{\star}<10^{-2.4}$) is $10^{-3.6}$, which is in excellent agreement with the $10^{-3.5}$ value inferred from stacking. In the sSFR-selected sample, however, there is a tail of galaxies with high gas fractions ($\gtrsim 0.1$) that do not seem to be observed. \citet{2026MNRAS.548ag375S} showed that COLIBRE predicts a downturn in the quenched fraction of galaxies above $M_{\star}\approx2\times10^{11}\,\mathrm{M}_{\odot}$ (their fig.~18) that these massive, H\,\textsc{i} gas-rich galaxies contribute to. This downturn may be in tension with the GAMA \citep{2013MNRAS.434..209B} and COSMOS/UltraVISTA \citep{2013ApJ...777...18M} surveys, but they note that (i) the observed quenched fraction becomes uncertain at these extremely high stellar masses and (ii) the tension disappears if the star formation rate is measured in a smaller $10\,\mathrm{kpc}$ aperture or if $0.3\,\mathrm{dex}$ lognormal scatter is added to the COLIBRE stellar masses. The relatively small number of FAST detections in the sample also means that the tail is poorly sampled in the observations, but we do find a related discrepancy with the larger ALFALFA sample (see Sec.~\ref{sec:tfr-btfr} below), so the lack of observed galaxies at $M_{\mathrm{HI}}/M_{\star}\gtrsim0.1$ here is likely not merely a finite sampling effect. Our finding that there may be a (small) population of massive quenched galaxies with H\,\textsc{i} gas fractions higher than is observed therefore hints at some mild disagreement with observations in this regime. On the other hand, most of the tail of higher H\,\textsc{i} gas fraction galaxies in the simulations have sSFR values slightly higher than the range sampled by the observations. Furthermore, the disagreement is only present in the sSFR-selected sample. The colour-selected sample agrees remarkably well with the observations. This suggests that uncertainties around stellar mass, star formation rate, and dust attenuation measurements in both the simulations and observations are important.

\subsubsection{The H\,\textsc{i} mass-size relation} \label{sec:mass-size}

The H\,\textsc{i} mass-size relation is a very tight scaling relation with a scatter of $0.07\,\mathrm{dex}$ and a power-law slope indistinguishable from $0.5$. This reflects the characteristic surface density of H\,\textsc{i} discs: these rarely reach azimuthally averaged surface densities in excess of $9\,\mathrm{M}_{\odot}\,\mathrm{pc}^{-2}$ \citep[$\approx 1.1 \times 10^{21}\,\mathrm{cm}^{-2}$;][]{2008AJ....136.2846B}, and the size used in the relation is defined as the diameter (or radius) where the H\,\textsc{i} surface density drops below $1\,\mathrm{M}_{\odot}\,\mathrm{pc}^{-2}$ ($\approx 1.2 \times 10^{20}\,\mathrm{cm}^{-2}$), so there is relatively little dynamic range available for the mean surface density to vary within this radius \citep[][reach a similar conclusion]{2026MNRAS.549ag762R}. This readily explains the slope, but the scatter need not \textit{a priori} be as tight as observed \citep[but see also][]{2019MNRAS.490...96S}.

\begin{figure}
  \includegraphics[width=\columnwidth]{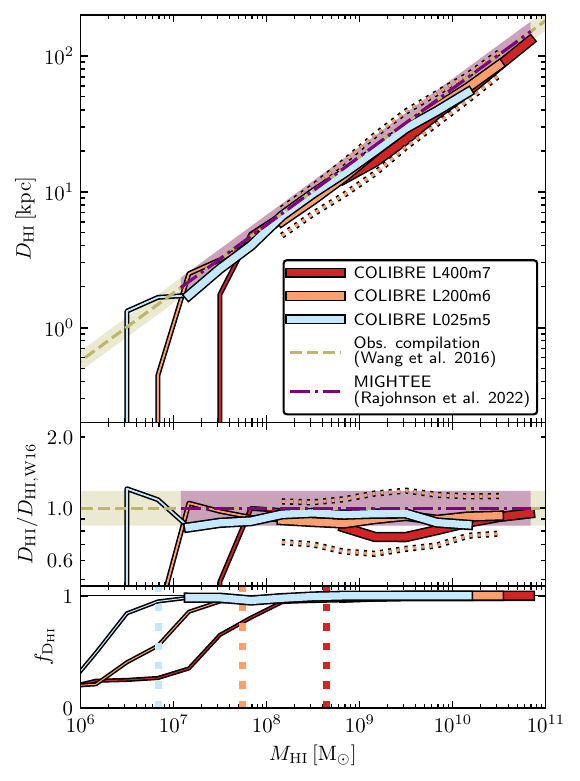}
  \caption{H\,\textsc{i} mass-size relation. $D_{\mathrm{HI}}$ is the diameter where the azimuthally-averaged H\,\textsc{i} surface density drops below $1\,\mathrm{M}_{\odot}\,\mathrm{pc}^{-2}$. The solid lines show median trends for the different COLIBRE resolution levels. The $16^{\mathrm{th}}$-$84^{\mathrm{th}}$~percentile scatter ($\approx 0.2\,\mathrm{dex}$) is shown with dotted lines only for the L200m6 simulation for clarity, it is similar in all $3$ simulations shown. Dashed and dash-dotted lines and shaded regions show fits to observational data from \citet{2016MNRAS.460.2143W} and \citet{2022MNRAS.512.2697R}, and scatter about them. Residuals with respect to the \citet{2016MNRAS.460.2143W} relation are shown in the second panel. The third panel shows the fraction of galaxies of given H\,\textsc{i} mass for which a size could be determined (see Sec.~\ref{sec:mass-size} for details). We plot the mass-size relation with thinner lines in bins including galaxies with $M_{\mathrm{HI}}\leq 30m_{\mathrm{g}}$; this limit is marked by the dotted lines. This panel shows that this limit ensures that nearly all galaxies have a measurable size. COLIBRE galaxies have H\,\textsc{i} that tends to be slightly too compact, although still within the very tight scatter observed, except for the L400m7 simulation which strays slightly further from the observed relation.}
  \label{fig:MHI-rHI}
\end{figure}

We show the mass-size relation for COLIBRE galaxies from the L025m5, L200m6 and L400m7 simulations in Fig.~\ref{fig:MHI-rHI}, compared with fits to samples of observed galaxies by \citet{2016MNRAS.460.2143W} and \citet{2022MNRAS.512.2697R} -- the two studies use independent data and find nearly identical relations and scatter. \citet{2022MNRAS.512.2697R} apply an inclination correction when measuring sizes, while \citet{2016MNRAS.460.2143W} use a compilation of pre-existing and new measurements of a mix of inclination-corrected and uncorrected sizes. Since the two measurements agree so closely we conclude that inclination effects do not strongly affect them. We estimate the H\,\textsc{i} sizes of COLIBRE galaxies in an approximate, but orientation-independent, way. We measure the H\,\textsc{i} mass contained in concentric spherical shells, then evaluate a surface density profile by dividing by the area of an annulus with the same inner and outer radii as each shell. This precludes the need to define a `disc plane'. We interpolate between the centres of the radial bins to estimate the outermost radius where the surface density drops below $1\,\mathrm{M}_{\odot}\,\mathrm{pc}^{-2}$. This approximation can fail badly for small radii (where the spherical shells have radii comparable to the disc thickness) or non-disc morphologies, but, since the characteristic size usually occurs at radii much larger than the disc thickness, the approximation is quite accurate. It systematically overestimates sizes by about $3$~per~cent (S.~Lee, priv. comm.), less even than the small scatter in the mass-size relation. We do not correct for this bias in Fig.~\ref{fig:MHI-rHI}.

Surface density profiles of simulated galaxies that have too few particles become shot-noise dominated and may (mostly by chance) exceed the threshold surface density nowhere. We assess the mass regime where this becomes important by measuring the fraction of galaxies that have an H\,\textsc{i} size according to the adopted definition as a function of H\,\textsc{i} mass, shown in the bottom panel of Fig.~\ref{fig:MHI-rHI}. We show the mass-size relation with thinner lines where galaxies have $M_{\mathrm{HI}}<30m_{\mathrm{g}}$ -- the bottom panel shows that this is a good lower limit that scales predictably with simulation resolution and avoids the shot-noise-dominated regime where sizes cannot be reliably measured.

The H\,\textsc{i} mass-size relation is very well reproduced in the COLIBRE simulations at all resolution levels; at worst the L400m7 simulation produces discs that are $25$~per~cent too small around $M_{\mathrm{HI}}=10^{9.5}\,\mathrm{M}_{\odot}$, while the higher resolution runs underpredict sizes by $\leq15$~per~cent. The scatter in the relation, shown only for the L200m6 run for clarity, is small ($0.2\,\mathrm{dex}$), though about twice that observed. A more detailed comparison with new size measurements from the WALLABY survey \citep{2020Ap&SS.365..118K} is in preparation by S.~Lee et al.

\subsubsection{The Tully-Fisher \& baryonic Tully-Fisher relations} \label{sec:tfr-btfr}

The Tully-Fisher \citep{1977A&A....54..661T,2001ApJ...550..212B,2017MNRAS.469.2387P,2019MNRAS.487.2924B} and baryonic Tully-Fisher \citep{2000ApJ...533L..99M,2019MNRAS.484.3267L,2021MNRAS.508.1195P,2023ApJ...950...87B} relations are scaling relations between the stellar \& neutral gas and total matter content of galaxies. The classical Tully-Fisher relation traces the correlation between the absolute magnitude (proxy for luminous mass) and inclination-corrected H\,\textsc{i} line width (proxy for dynamical mass). It was later found that including H\,\textsc{i} gas in the visible matter budget results in a much tighter correlation. A common parametrization for the `baryonic mass' in this context is $M_{\mathrm{b}}=M_{\star}+C_{\mathrm{heavy}}M_{\mathrm{HI}}$ where $C_{\mathrm{heavy}}$ is a correction to the atomic gas mass accounting for helium and heavier elements. Commonly assumed values for $C_{\mathrm{heavy}}$ are $4/3$ and $1.4$ -- the former assumes a near-primordial gas composition while the latter assumes more helium and metal enrichment. This choice has negligible influence on the results presented in this section; we adopt $C_{\mathrm{heavy}}=1.4$. The contributions from molecular and ionized gas are usually omitted. We follow this convention, but see \citet{2020RNAAS...4...45M} for a discussion of the influence of including molecular gas on the baryonic Tully-Fisher relation.

Our main aim in this section is to compare the statistics of the joint distributions of stellar/baryonic masses and H\,\textsc{i} line widths of COLIBRE galaxies against those observed. Contemporary studies of both relations often use heterogeneous galaxy samples with intricate selection criteria. Reproducing these sufficiently accurately to enable a fair comparison is challenging \citep[see][who are working towards such an effort, and their sec.~5.2 for a discussion of remaining hurdles]{2025A&A...696A..52B} and risks putting most of the emphasis in interpretation on the influence of various selection effects and biases. We therefore instead choose to compare against the ALFALFA galaxy catalogue, whose selection function we can accurately reproduce, cross-matched against the Sloan Digital Sky Survey \citep[SDSS;][]{2020AJ....160..271D} to supplement it with stellar mass measurements. The two surveys overlap only over about 2/3 of the ALFALFA survey area.

We use the same mock galaxy catalogues as in Sec.~\ref{sec:himf-hiwf}, but reduce the volume to account for the smaller volume that overlaps with the SDSS. The Tully-Fisher and baryonic Tully-Fisher relations are usually defined using line widths (or rotation speeds) corrected for inclination to obtain a more direct proxy for galaxies' dynamical masses. These corrections require notoriously uncertain geometric assumptions, especially for inclination. Uncertainties are further exacerbated when inclinations estimated from optical photometry are assumed to also hold for the H\,\textsc{i} gas, as for ALFALFA. On the other hand, our mock survey approach self-consistently includes the effect of inclination in the mock spectra from which we measure line widths. We therefore choose to compare measured line widths (without inclination correction) to remove a potential source of systematic errors.

\begin{figure*}
  \includegraphics[width=\textwidth]{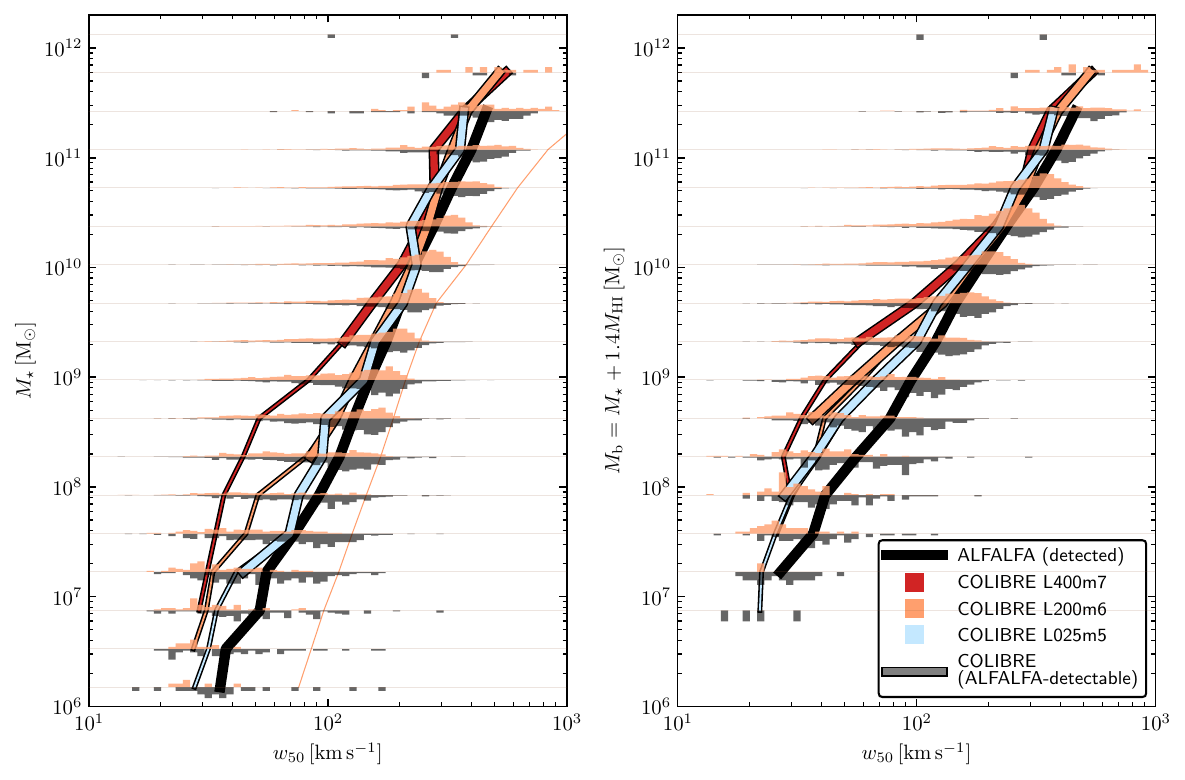}
  \caption{The Tully-Fisher (stellar mass $M_{\star}$ vs. H\,\textsc{i} line width $w_{50}$) and baryonic Tully-Fisher (`baryonic mass' $M_{\mathrm{b}}=M_{\star}+1.4M_{\mathrm{HI}}$ vs. $w_{50}$) relations. \textit{Left panel:} The Tully-Fisher relation. Uncertain inclination corrections to $w_{50}$ are not applied (neither to ALFALFA nor COLIBRE data), instead inclination effects are forward-modelled for the simulated galaxies, which is less uncertain. Heavy lines show the median line width in bins of stellar mass for ALFALFA galaxies (black) and COLIBRE galaxies as labelled, using only galaxies with at least $100$ stellar particles. The thinner extensions of these lines are instead computed using all galaxies. Histograms showing the full distributions for ALFALFA and the COLIBRE L200m6 simulation in the same stellar mass bins are overlaid. Each pair of histograms (i.e. within a stellar mass bin) shares the same (arbitrary) vertical scale so that relative counts can be compared visually; the normalisation is adjusted across stellar mass bins for clarity. The thin line shows the upper limit of twice the maximum circular velocity ($2v_{\mathrm{max}}$) for simulated galaxies in the same stellar mass bins -- line widths almost never exceed this limit. \textit{Right panel:} As the left panel but replacing stellar mass with `baryonic mass', $M_{\mathrm{b}}=M_{\star}+1.4M_{\mathrm{HI}}$.}
  \label{fig:tfr-btfr}
\end{figure*}

The $M_{\star}$-$w_{50}$ Tully-Fisher relation and $M_{\mathrm{b}}$-$w_{50}$ baryonic Tully-Fisher relation for COLIBRE galaxies are compared against ALFALFA measurements\footnote{Stellar masses are estimated as $\log_{10}\left(M_{\star}/\mathrm{M}_{\odot}\right)=1.052\log_{10}\left(M_{\star}/\mathrm{M}_{\odot}\right)_{\mathrm{Taylor}}-0.369$, see eq.~(3) in \citet{2020AJ....160..271D}. $\log_{10}\left(M_{\mathrm{Taylor}}/\mathrm{M}_{\odot}\right)$, in turn, is a simple mass estimator based on optical magnitude and luminosity, see eq.~2 in \citep{2020AJ....160..271D}. These stellar masses assume a \citep{2003PASP..115..763C} initial mass function, the same that is assumed in COLIBRE.} in the left and right panels of Fig.~\ref{fig:tfr-btfr}, respectively. The heavy lines show the median trends for $w_{50}$ in bins of stellar mass. COLIBRE galaxies are offset to lower line widths in the median, although the agreement improves with increasing resolution (the curve for L025m5 is somewhat noisier than the others due to the small box size). The agreement is quite good for the L200m6 and L025m5 simulations where most of the observed galaxy population resides ($10^{9}<M_{\star}/\mathrm{M}_{\odot}<10^{10.5}$ or $10^{9.5}<M_{\mathrm{b}}/\mathrm{M}_{\odot}<10^{11}$), but more nuanced impressions can be obtained by inspecting the full distribution of $w_{50}$ in each mass bin rather than only the median.

These distributions are overlaid on the figure as histograms. We break the main features down into three regimes at low ($\lesssim 5\times10^{8}\,\mathrm{M}_{\odot}$), intermediate ($5\times 10^{8}$-$10^{11}\,\mathrm{M}_{\odot}$) and high ($\gtrsim 10^{11}\,\mathrm{M}_{\odot}$) stellar masses (or roughly equivalent boundaries at $M_{\mathrm{b}} \approx 10^{9.5}\,\mathrm{M}_{\odot}$ and $10^{11}\,\mathrm{M}_{\odot}$). The best agreement between the L200m6 COLIBRE simulation and ALFALFA is in the intermediate mass regime. Inspecting the distributions, their modes align essentially perfectly in this mass range, with the offset in the median being driven mostly by the tails: the distribution for the simulations has a heavier tail to low-$w_{50}$ while ALFALFA has a longer tail at the high-$w_{50}$ end. There are also differences in the number counts (i.e. the integrals of the histograms) for COLIBRE and ALFALFA in each mass bin, but these reflect the same discrepancies in galaxy counts already discussed in Sec.~\ref{sec:himf-hiwf}. The presence of extreme outliers towards large $w_{50}$ values in the ALFALFA catalogue suggests that at least part of the high-$w_{50}$ tail of observed galaxies is probably due to measurement errors (e.g. we do not mimic the ALFALFA spectral noise properties in our mocks) or misassociations of optical and H\,\textsc{i} sources. For example, a few galaxies with $M_{\star} \approx 10^{9}\,\mathrm{M}_{\odot}$ have $w_{50}>400\,\mathrm{km}\,\mathrm{s}^{-1}$ which seems implausible on physical grounds. We further illustrate this with the thin line on the figure. This shows the $99.9^{\mathrm{th}}$ percentile of the distribution of twice the maximum circular velocity ($2v_{\mathrm{max}}$) in each mass bin. This is approximately the line width expected for a fully rotationally-supported H\,\textsc{i} disc that is large enough to reach the radius where $v_{\mathrm{max}}$ is reached and seen edge-on. COLIBRE galaxies never exceed this limit by more than a few per~cent -- at higher and lower masses they do not reach it as the H\,\textsc{i} discs are less extended, less rotationally-supported, or both. The thin line is a reflection of the stellar-to-halo mass relation, so if the simulations are biased away from the true relation an offset could arise, but since the peaks of the distributions agree so well this interpretation seems less likely. \citet{2026MNRAS.548ag375S} also showed that the stellar-to-halo mass relation for COLIBRE galaxies agrees extremely well with observational constraints.

At lower stellar (or baryonic) masses the simulations under-produce mock survey detections, and those simulated galaxies that are detectable have narrower H\,\textsc{i} lines on average than observed. These discrepancies are related because the survey is more sensitive to narrow lines, so as the detection rate drops a bias towards narrower lines is expected. Simulated galaxies having lower H\,\textsc{i} masses than observed at fixed $M_{\star}$ in this regime can partially explain this discrepancy. However, as the bias is also present in the baryonic Tully-Fisher relation, and these low-mass galaxies predominantly have baryonic masses dominated by the contribution from H\,\textsc{i}, other effects must also be at play. Fig.~\ref{fig:Mstar-MHI} showed that in the COLIBRE L200m6 model the H\,\textsc{i} mass distribution becomes strongly bimodal below $M_{\star} \approx 10^{8}\,\mathrm{M}_{\odot}$. In this subset of the galaxy population, those galaxies that do contain H\,\textsc{i} have it in great abundance, and these are the galaxies that are detected and contribute to the (baryonic) Tully-Fisher relation, but the fraction of such galaxies drops precipitously towards lower stellar masses. Whether an individual galaxy retains its H\,\textsc{i} is a strongly non-linear function of its history, so some difficulty capturing this exactly in a simulation is understandable. Furthermore, this regime pushes the boundaries imposed by the numerical resolution of the simulations. This is clear from the median curves -- these no longer overlap where the lines become thinner, i.e. where galaxies have less than $100$ stellar particles. We have checked that the shapes of the histograms for the L025m5 simulation agree much better with the ALFALFA observations down to $M_{\star} \approx 10^{7}\,\mathrm{M}_{\odot}$ (not shown), although this simulation tends to have H\,\textsc{i} masses biased high across the entire galaxy population as discussed in Sec.~\ref{sec:himf-hiwf}, which also leads to poorer agreement at the low-mass end of the baryonic Tully-Fisher relation.

Finally, at the high-mass end of both relations shown in Fig.~\ref{fig:tfr-btfr} (for these galaxies $M_{\mathrm{b}}\approx M_{\star}$) we find a population of simulated galaxies that do not appear in ALFALFA observations. COLIBRE does accurately reproduce the abundance of galaxies of these stellar (or baryonic) masses much better than this figure suggests \citep[see Sec.~\ref{sec:himf-hiwf} and][sec.~7.1.1]{2026MNRAS.548ag375S} -- the discrepancy is rather because they do not appear in the H\,\textsc{i}-selected ALFALFA catalogue, but do appear in our mock catalogue. In Sec.~\ref{sec:massive-quenched} we noted overall good agreement in the H\,\textsc{i} gas content of massive, quiescent galaxies, but noted the presence of a small tail of simulated massive galaxies with $M_{\mathrm{HI}}/M_{\star}>10^{-1}$ that is not observed. On closer inspection, these galaxies are preferentially found at the upper extreme of the stellar mass function ($M_{\star}\gtrsim 5\times 10^{11}\,\mathrm{M}_{\odot}$), giving them H\,\textsc{i} masses of $\approx 10^{10}\,\mathrm{M}_{\odot}$ -- enough to be detected anywhere in the ALFALFA survey volume. The simulated galaxies with $M_{\star}\approx M_{\mathrm{b}} \approx 10^{12}\,\mathrm{M}_{\odot}$ that are detected in our mock survey are the same as those making up the high-gas fraction tail in Fig.~\ref{fig:massive-quenched} (see discussion in Sec.~\ref{sec:massive-quenched}).

In summary, COLIBRE reproduces the ALFALFA Tully-Fisher and baryonic Tully-Fisher relation in the median and full shape of the scatter where galaxies are resolved by more than $\approx 100$ stellar particles, except at the extreme high-mass end ($M_\star\gtrsim2\times10^{11}\,\mathrm{M}_{\odot}$).

\subsubsection{H\,\textsc{i} surface density profiles} \label{sec:SigmaHI}

We have so far discussed integral quantities. We now turn to a first look at radial H\,\textsc{i} surface density profiles, which offer a straightforward leading-order characterisation of the spatial distribution of H\,\textsc{i} in galaxies. We already established in Sec.~\ref{sec:mass-size} that H\,\textsc{i} discs in COLIBRE are slightly (about $10$~per~cent) too compact when size is characterized by the radius where the surface density drops below $1\,\mathrm{M}_{\odot}\,\mathrm{pc}^{-2}$. In this section we instead focus mostly on the density regime well below this threshold.

\citet{2025ApJ...980...25W} recently published very deep H\,\textsc{i} surface density profiles\footnote{See also \citet{2026A&A...712A..74V} for a complementary approach using spectral stacking of interferometric observations from the MHONGOOSE survey.} for a sample of $35$ nearby galaxies observed as part of the FEASTS\footnote{FAST Extended Atlas of Selected Targets Survey.} survey on FAST. We adopt their sample as a comparison benchmark.

\begin{figure*}
\includegraphics[width=\textwidth]{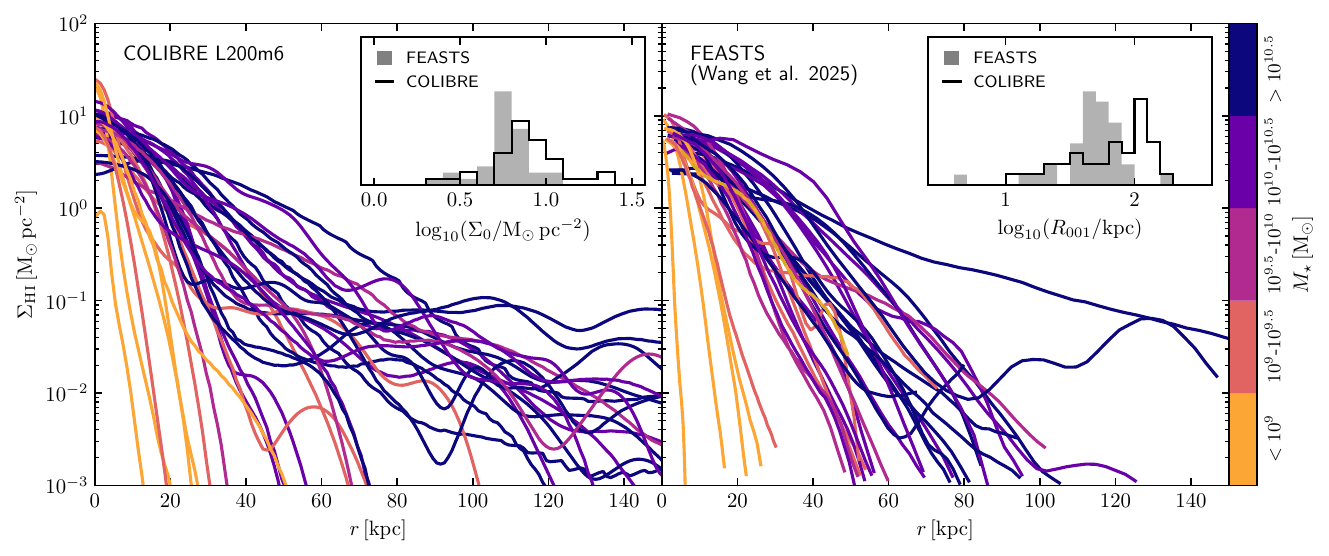}
\caption{H\,\textsc{i} surface density profiles for a sample of galaxies matched in stellar mass, H\,\textsc{i} mass, distance, inclination and noise level to the FEASTS survey sample of \citet[][fig.~2]{2025ApJ...980...25W}. \textit{Left panel:} H\,\textsc{i} surface density as a function of radius for the matched sample from the L200m6 simulation. Galaxies are first mock-observed, and the resulting data are then analysed using source finding and profile extraction methods closely mimicking those used for FEASTS (see Sec.~\ref{sec:SigmaHI} and Appendix~\ref{app:mocks-feasts}). The lines are coloured by galaxy stellar mass in $5$ bins, as labelled. \textit{Right panel:} Similar to left panel, but for the galaxies observed in the FEASTS survey. \textit{Inset panels:} The left inset shows the distribution of central ($r=0$) H\,\textsc{i} surface densities $\Sigma_{0}$ for the observed (filled histogram) and simulated (open histogram) galaxies. The simulated galaxies are slightly offset ($\approx 0.1\,\mathrm{dex}$) to higher central surface densities. The highest-density outliers are among the galaxies with the lowest stellar masses in the sample ($M_{\star}\lesssim10^{9.5}\,\mathrm{M}_{\odot}$). The right inset shows the distribution of $R_{001}$, the radius where the profile crosses $\Sigma_{\mathrm{HI}}=0.01\,\mathrm{M}_{\odot}\,\mathrm{pc}^{-2}$, with the same histogram styles. Several simulated galaxies have profiles more extended than those observed. These are predominantly among the most massive galaxies in the sample ($M_{\star}\gtrsim 10^{10}\,\mathrm{M}_{\odot}$). This seems to be related to the presence of the discrete H\,\textsc{i} clouds, especially around more massive galaxies, that do not seem to be as common in observed systems (Sec.~\ref{sec:cgm-clouds}).}
\label{fig:sig-profiles}
\end{figure*}

We construct mock-observed surface density profiles attempting to closely mimic the FEASTS observations and methodology for measuring profiles. Details are provided in Appendix~\ref{app:mocks-feasts}; here we provide a brief summary. For each of the $35$ galaxies in the FEASTS sample, we select one central galaxy from the COLIBRE L200m6 simulation that closely matches its stellar mass and its H\,\textsc{i} mass. We place the simulated galaxy (along with its companions and environment out to a radius of $250\,\mathrm{kpc}$) at the same distance and approximately the same inclination as the observed galaxy from a notional observer and produce a data cube that matches the spatial \& spectral resolution and sensitivity of the FEASTS observations. We visually inspect the resulting data cubes and note that $9$ galaxies in our sample of simulated galaxies show clear signs of interaction with a close companion. \citet{2025ApJ...980...25W} identify $11$ likely interacting galaxies in their sample. As this classification is subjective and interactions come in many forms, we do not attempt to match interacting galaxies one-to-one but simply conclude that the simulated and observed samples are not grossly different in this regard. We run the \textsc{SoFiA-2} source finding algorithm on each simulated data cube using the same configuration as was used for FEASTS to extract the emission corresponding to the most massive H\,\textsc{i} source and measure its surface density profile, applying an inclination correction analogous to that used by \citet{2025ApJ...980...25W}.

The resulting profiles are shown in the left panel of Fig.~\ref{fig:sig-profiles}. We reproduce the corresponding $35$ observed surface density profiles from FEASTS in the right panel. Many profiles are very similar to those observed, but some differences are also apparent.

First, many profiles have strong outer ($r\gtrsim 50\,\mathrm{kpc}$) fluctuations in surface density and extend to large radii ($r>100\,\mathrm{kpc}$); only $2$ observed profiles have emission above $10^{-2}\,\mathrm{M}_{\odot}\,\mathrm{pc}^{-2}$ beyond $100\,\mathrm{kpc}$ compared to $15$ in the simulation sample. This is visualised in the inset in the right panel that shows the distribution of radii $R_{001}$ where the profiles first drop below $\Sigma_{\mathrm{HI}}=0.01\,\mathrm{M}_{\odot}\,\mathrm{pc}^{-2}$ for the simulated and observed samples. Much of the observed distribution is well-reproduced, but there is a subset of simulated galaxies that have $R_{001}\gtrsim 100\,\mathrm{kpc}$. The emission at large radii seems to be sourced by a population of H\,\textsc{i} clouds surrounding more massive galaxies, similar to those noticed in Sec.~\ref{sec:visual-impressions}. We explore these further in Sec.~\ref{sec:cgm-clouds}, where we will show that they seem to be more abundant than is observed. Indeed, the profiles with extended emission above $\Sigma_{\mathrm{HI}}=10^{-2}\,\mathrm{M}_\mathrm{\odot}\,\mathrm{pc}^{-2}$ are all amongst the most massive galaxies in the sample ($M_{\star}\gtrsim10^{10}\,\mathrm{M}_{\odot}$, darker coloured lines).

The abundance of the circumgalactic clouds contributing the emission at large radii is sensitive to the uncertain radiation shielding length assumed in the model. The abundance of such clouds is reduced by a factor of $\approx2$ when the shielding length is halved, whic is easily within the uncertainties given that the shielding length is assumed to be equal to the Jeans' length. The definition of the Jeans' length is uncertain at the factor of two level, and its use is only appropriate for self-gravitating clouds. The shielding length should be reduced further if the clouds are confined by external pressure rather than by their self-gravity. A stronger $z=0$ metagalactic X-ray/UV radiation field, or accounting for the fact that it may be stronger than the cosmic average in the vicinity of massive galaxies, may also help to suppress the formation of the clouds. These issues are discussed further in Sec.~\ref{sec:cgm-clouds} and Appendix~\ref{app:vars}, where we also show that the abundance and H\,\textsc{i} mass in the circumgalactic clouds decreases by a factor of up to $3$ and $20$ (but less above $M_{\mathrm{HI}}>2\times10^{9}\,\mathrm{M}_{\odot}$), respectively, if the mass resolution is increased by a factor of $8$.

The lower mass galaxies ($M_{\star}\lesssim10^{9.5}\,\mathrm{M}_{\odot}$) in the sample generally have smooth, steeply declining profiles very similar to those of their observed counterparts.

We also briefly comment on the densities measured at the centres of galaxies. The FAST beam has a full width at half maximum of $3.24\,\mathrm{arcmin}$, corresponding to $3$-$23\,\mathrm{kpc}$ at the distances of galaxies in the FEASTS sample, so beam smoothing will often cause the apparent surface density to be lower than the true surface density. However, since our matched sample approach reproduces the distance distrbution and the beam smoothing by construction, a like-for-like comparison is possible. The simulated galaxies have a similar distribution of central H\,\textsc{i} surface densities $\Sigma_{0}$ to that observed, but shifted to slightly higher densities (by $\approx0.1\,\mathrm{dex}$). There are also a few outliers at $1.2\lesssim\log_{10}(\Sigma_{0}/\mathrm{M}_{\odot}\,\mathrm{pc}^{-2})\lesssim1.4$ that have no observed counterparts; these are all among the galaxies with the lowest stellar masses in the sample ($M_{\star}\lesssim10^{9.5}\,\mathrm{M}_{\odot}$), although other galaxies with similar masses and central densities very similar to those of their observed counterparts are also present in the sample.

This initial comparison of COLIBRE H\,\textsc{i} surface density profiles with observations is only a very simple first look at the internal H\,\textsc{i} structure of galaxies resolved in the COLIBRE simulations. Fully understanding the emergent H\,\textsc{i} morphology and kinematics of COLIBRE galaxies offers many interesting avenues for future study.

\subsubsection{Central holes} \label{sec:central-holes}

\begin{figure}
  \includegraphics[width=\columnwidth]{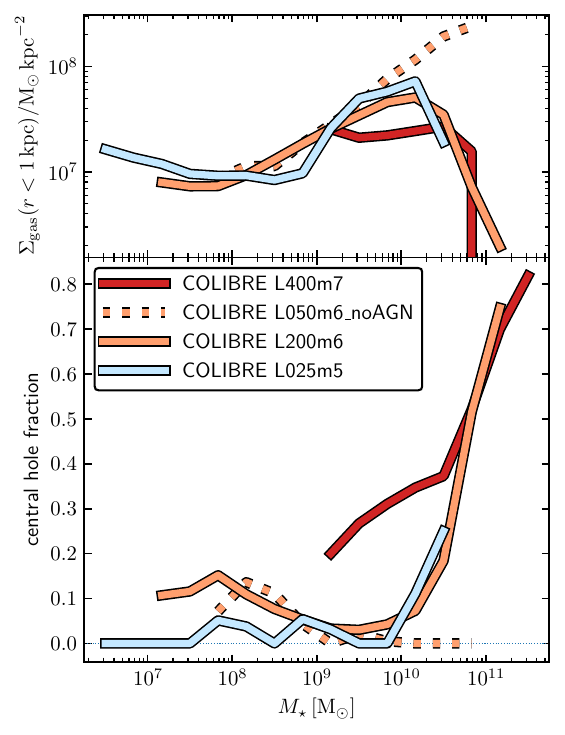}
  \caption{Assessment of the frequency of central gas `holes' in disc galaxies as a function of stellar mass. `Disc' galaxies are selected by gas fraction and kinematics (see Sec.~\ref{sec:central-holes}). \textit{Upper panel:} Median central gas surface density estimated as gas mass within a spherical aperture of $1\,\mathrm{kpc}$ radius divided by the surface area of a disc inscribed within it. The sharp drop towards high stellar masses (solid lines) hints at the scale where central holes become common. The drop is not present in a variant run without AGN feedback (L025m6\_noAGN; dotted line), pointing to an AGN-feedback origin for the holes. \textit{Lower panel:} Fraction of disc galaxies hosting a central gas hole, defined as a drop by a factor of $2$ in mean gas surface density within $1\,\mathrm{kpc}$ compared to within $10\,\mathrm{kpc}$ (see Sec.~\ref{sec:central-holes}). Holes become common above $M_{\star}=10^{10}\,\mathrm{M}_{\odot}$. In both panels, only bins (of $0.3\,\mathrm{dex}$ width) containing at least $10$ galaxies are visualised.}
  \label{fig:holes}
\end{figure}

We now investigate the incidence of the central holes in gas discs noted in Figs.~\ref{fig:visual-impression-discs} \& \ref{fig:visual-impression-groups}. We estimate this as the fraction of disc galaxies where the average gas surface density within $1\,\mathrm{kpc}$ is less than half the average gas surface density within $10\,\mathrm{kpc}$. We define `disc galaxies' in this context as those with gas fractions $M_{\mathrm{HI}}/M_{\star}>0.1$ and kinematic disc-to-total mass fractions $>0.3$ for both the stellar and gas components. The disc-to-total mass fraction is the fraction of material remaining once twice the mass that is counter-rotating with respect to the total angular momentum axis has been removed. We also only include galaxies with at least $1000$ gas particles within a $50\,\mathrm{kpc}$ spherical aperture to ensure that the disc morphology is adequately sampled to search for a hole. The gas surface densities are approximated as the gas masses in spherical apertures divided by the area of discs with the same radii. We use the total gas mass (not H\,\textsc{i} mass) for this purpose to emphasize that the holes are not merely due to a deficit of H\,\textsc{i} where, for instance, the molecular fraction rises, but are genuine evacuated holes. The fraction of discs with a central hole according to this definition is visualised as a function of stellar mass in Fig.~\ref{fig:holes}.

The upper panel of the figure shows the (approximate) gas surface density in the central $1\,\mathrm{kpc}$ as a function of galaxy stellar mass. This reveals that above $M_{\star}\approx 10^{10}\,\mathrm{M}_{\odot}$ the median central surface density drops precipitously in the L200m6 model (solid orange line). At lower (L200m7) and higher (L025m5) resolution a similar decline occurs at slightly higher and lower stellar masses, respectively. We also show results for a small volume simulated at the intermediate resolution level with a model variation with AGN feedback switched off (L025m6\_noAGN; dotted line). In this case, the median central surface densities do not drop at high stellar masses, confirming the AGN-driven origin of the holes. The lower panel of the figure shows the fraction of disc galaxies with a central hole in their gas disc (according to the definition outlined above). The drop in central surface density seen in the upper panel is reflected in a rise in the occurence rate of holes. Such holes are common in galaxies with $M_{\star}>10^{10}\,\mathrm{M}_{\odot}$ in all models (solid lines), and are ubiquitous when $M_{\star}>10^{11}\,\mathrm{M}_{\odot}$, although their precise frequency is, unsurprisingly, resolution-dependent, with holes appearing more frequently at fixed stellar mass in lower resolution simulations. For instance, the higher energy per AGN feedback energy injection event invoked at the lower (m7) resolution likely promotes the formation of gas holes.

Although galaxies with central gas holes are observed \citep[e.g.][and notably NGC~2775, \citealp{2001AJ....121.1336H}]{1999ApJS..124..403S,2019MNRAS.483.2398M,2024A&A...689A.347G}, those in the COLIBRE simulations seem to be larger (up to several kiloparsecs) and more common than in the observed galaxy population. Many observed holes also turn out to be in one tracer (e.g. H\,\textsc{i}) but `filled in' by gas in a different phase \citep[e.g. H$_{2}$,][]{2002ApJ...569..157W}, whereas the holes in COLIBRE discs are more often completely evacuated of gas. The dependence on AGN feedback (see also Thompson et al., in preparation) suggests that the properties and statistics of central gas holes might provide an interesting constraint on these models, though we caution that both are subgrid implementations -- on these small scales constraining the AGN physics may be too optimistic, but constraining its effect on the properties of the gas disc may be achievable.

\subsubsection{Circumgalactic \texorpdfstring{H\,\textsc{i}}{HI} clouds} \label{sec:cgm-clouds}

To estimate the abundance of H\,\textsc{i} clouds surrounding COLIBRE galaxies like those noted in Figs.~\ref{fig:visual-impression-discs} \& \ref{fig:visual-impression-groups}, we make use of a simple friends-of-friends analysis. Using the \textsc{soap} utility, we extract all particles identified as gravitationally bound to that galaxy (and not to its satellites) by the \textsc{hbt-herons} finder that have an H\,\textsc{i} fraction (relative to total hydrogen) of $>0.5$. Since the H\,\textsc{i} fraction distribution is strongly bimodal, this cleanly selects H\,\textsc{i}-rich particles. We then run a friends-of-friends group search on these particles with a linking length of $3\,\mathrm{kpc}$, chosen to be slightly larger than the mean spacing between H\,\textsc{i}-rich particles in the discs of galaxies in the L400m7 simulation (corresponding to a characteristic surface density $\approx 2\,\mathrm{M}_{\odot}\,\mathrm{pc}^{-2}$). This may occasionally link nearby clouds to the disc or to each other, but is sufficient for our purpose of informing an initial impression. We label each group a `cloud', except the most massive one that we assume is the `primary' H\,\textsc{i} associated with the galaxy (e.g. the disc). We retain all groups, including those made up of a single particle.

\begin{figure}
\includegraphics[width=\columnwidth]{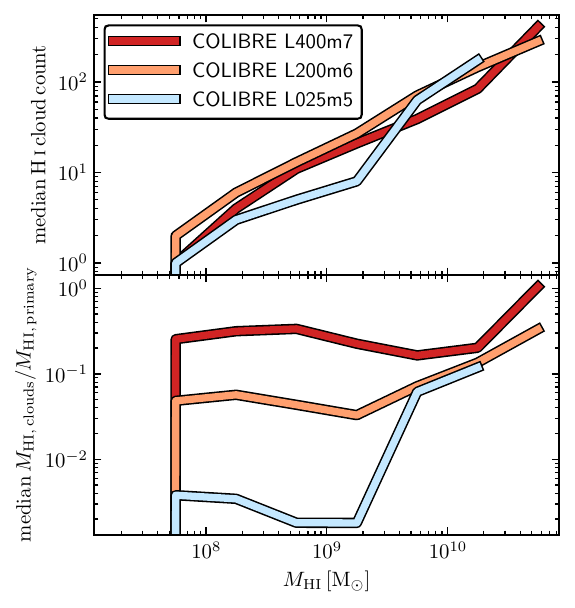}
\caption{Assessment of the prevalence of circumgalactic H\,\textsc{i} clouds as a function of total H\,\textsc{i} mass. \textit{Upper panel:} The median number of clouds per galaxy identified by a friends-of-friends search linking groups of H\,\textsc{i}-rich particles (see Sec.~\ref{sec:cgm-clouds}). Clouds begin to appear above $M_{\mathrm{HI}}=5 \times 10^{7}\,\mathrm{M}_{\odot}$ and their number count is converged within a factor of $\approx3$ across the three resolution levels. \textit{Lower panel:} The ratio of H\,\textsc{i} mass in clouds relative to the mass of the primary H\,\textsc{i} component associated with the galaxy. At lower resolution there is more mass in clouds at fixed total H\,\textsc{i} mass, especially at $M_{\mathrm{HI}}<10^{9.5}\,\mathrm{M}_{\odot}$ where differences exceed $1\,\mathrm{dex}$. Since the cloud number counts (upper panel) and H\,\textsc{i} masses of the primary component (not shown) converge across resolution levels to within a factor of $\lesssim 3$, this implies that the individual clouds have larger H\,\textsc{i} masses at lower resolution.}
\label{fig:clouds}
\end{figure}

We explored both the number of clouds and their cumulative mass relative to the mass of the `primary' component as a function of halo, stellar and H\,\textsc{i} mass (either total or of the `primary' component). Of these, the relations with total H\,\textsc{i} mass have the least scatter at fixed mass (scatter not shown), so we show the trends as a function of $M_{\mathrm{HI}}$ in Fig.~\ref{fig:clouds}. Clouds begin to appear above $M_{\mathrm{HI}}=5\times10^{7}\,\mathrm{M}_{\odot}$ at all resolution levels and the number of clouds scales approximately as a power law with total H\,\textsc{i} mass above this. While the number of clouds changes by at most about a factor of $3$ across the entire measurable mass range at fixed mass when the numerical resolution is increased or decreased by a factor of $8$, the total H\,\textsc{i} mass contained in the clouds relative to the mass of the `primary' component is more resolution-dependent. The mass in clouds drops by up to a factor of $20$ at low masses ($M_{\mathrm{HI}}\leq2\times10^{9}\,\mathrm{M}_{\odot}$) when the numerical resolution is increased by a factor of $8$, but is within a factor of $\approx 3$ at higher H\,\textsc{i} masses. We have checked that the H\,\textsc{i} mass of the `primary' component is well converged as a function of total H\,\textsc{i} mass (not shown), so the lower panel of Fig.~\ref{fig:clouds} reveals that at low H\,\textsc{i} masses (corresponding to $\lesssim 1000m_{\mathrm{g}}$ of H\,\textsc{i} in total, where $m_{\mathrm{g}}$ is the initial average mass of a simulation gas particle) the H\,\textsc{i} masses of the individual clouds are higher at lower resolution such that the H\,\textsc{i} mass contained in the cloud population is a strong function of resolution; at higher masses ($M_\mathrm{HI}\gtrsim10^{10}\,\mathrm{M}_{\odot}$) it is better converged.

Abundant populations of H\,\textsc{i} clouds similar to the examples seen in Figs.~\ref{fig:visual-impression-discs} \& \ref{fig:visual-impression-groups} seem to be ubiquitous around massive, H\,\textsc{i}-rich galaxies in COLIBRE. This seems to be contrary to observations: although H\,\textsc{i} emission is observed outside of the dense discs of galaxies \citep[see][for a recent example]{2026arXiv260819186R}, and whether this gas is in the form of a diffuse medium or compact clouds is still debated, they are constrained to have lower column densities than the clouds found in the simulations.

This is apparent from Figs.~\ref{fig:visual-impression-discs} \& \ref{fig:visual-impression-groups}: the resolution and sensitivity of those mock observations is matched to the MHONGOOSE survey, so similar clouds should be visible in data from that survey if they existed, yet the majority of MHONGOOSE fields contain very few detectable H\,\textsc{i} clouds outside of the primary survey targets and their satellite galaxies. We have visually counted clouds in the zeroth moment maps included in the `atlas' pages produced by the MHONGOOSE team\footnote{Available at \url{mhongoose.astron.nl/Public-Data-Release.html}.} \citep[see also][]{2025A&A...697A..86M}. The abundance of clouds does increase with H\,\textsc{i} mass as predicted by the simulations: below $M_{\mathrm{HI}}=10^{8.5}\,\mathrm{M}_{\odot}$ only NGC~1705 hosts a single visible cloud, while in the range $10^{8.5}<M_{\mathrm{HI}}/\mathrm{M}_{\odot}<10^{9.6}$ many galaxies have $3$-$4$ clouds (e.g. IC~1954, NGC~4781, NGC~1744 and others, with a mean count of $2$ clouds in this mass range). The most massive galaxies in the sample ($10^{9.8}<M_{\mathrm{HI}}/\mathrm{M}_{\odot}<10^{10.3}$) host up to $10$ clouds each (e.g. UGCA~250, NGC~1566, mean count of $6$ clouds). Although not all of the clouds identified by our friends-of-friends analysis are expected to be visible at the depth of the MHONGOOSE survey, the images in Figs.~\ref{fig:visual-impression-discs} \& \ref{fig:visual-impression-groups} show that for these massive galaxies the simulations predict that many more should be visible: there are about $90$ in field~A, and over $200$ each in fields~C \& D ($10^{9.9}<M_\mathrm{HI}/\mathrm{M}_{\odot}<10^{10.3}$).

There are two relatively minor caveats to this comparison. First, we have not mimicked the effects of primary beam attenuation in the images in Figs.~\ref{fig:visual-impression-discs} \& \ref{fig:visual-impression-groups}. This could make some clouds at larger radii invisible (the MeerKAT primary beam is $\approx 1^\circ$ at FWHM). However, there are still more H\,\textsc{i} clouds than are observed in the simulations even when only including those at smaller radii, and even the more sensitive and less spatially resolved reduction of the MHONGOOSE data (we mimic the highest-resolution, lowest-sensitivity reduction, \texttt{r00\_t00}) do not detect additional clouds. Second, we have not attempted to exclude faint satellites when counting clouds in the MHONGOOSE images, but these would be excluded from our friends-of-friends counts as they only include gas bound to the central galaxy and not its satellites. This would tend to increase the apparent discrepancy. It therefore seems that circumgalactic H\,\textsc{i} clouds are much more numerous in the simulations than is observed in the MHONGOOSE survey.

Another line of evidence comes from spectral stacking. Using this technique, again using the MHONGOOSE survey data, \citet{2025A&A...693A..97V} find no evidence of extended emission from H\,\textsc{i} at a spatially-averaged column density of $\gtrsim10^{17}\,\mathrm{cm}^{-2}$ (individual clouds would be invisible to this technique but, were they present, their cumulative emission might have been detected this way, although we caution that the spectral stacking method is best suited to searching for a low-density extension of the H\,\textsc{i} disc). The related but complementary comparison with observations from the FEASTS survey in Sec.~\ref{sec:SigmaHI} above similarly suggests that less H\,\textsc{i} is present in the circumgalactic medium of massive FEASTS galaxies than around their counterparts in COLIBRE, although those observations lack the spatial resolution to resolve individual clouds.

Abundant circumgalactic H\,\textsc{i} clouds are implied by Ly$\alpha$ emission observed around galaxies at higher redshifts, and the L025m5 COLIBRE simulation quantitatively reproduces this at $z\approx 3.5$ \citep{2026arXiv261001400L}, so the discrepancy seems to be time-dependent.

Since the $z=0$ H\,\textsc{i} clouds in the simulations seem to be absent from observations, we consider possible origins of this discrepancy. The abundance and H\,\textsc{i} mass in clouds is not very sensitive to assumptions around the interstellar radiation field intensity (they are more likely regulated by the metagalactic UV/X-ray background), but does depend on assumptions around the shielding length for ionizing radiation. Halving the shielding length leads to a factor of $\approx2$ change in both the number and total mass contained in clouds across most of the mass range where we can measure them. Further details on the dependence on these assumptions is elaborated in Appendix~\ref{app:vars} (Fig.~\ref{fig:clouds-mv}). Since the clouds are external pressure-confined, rather than self-gravitating, the shielding lengths inside them are by definition shorter than the Jeans' length used to approximate them, potentially by a large factor. This could lead to producing too many circumgalactic H\,\textsc{i} clouds. Alternatively, or in addition, since the clouds are likely regulated by the UV/X-ray background, they are expected to be sensitive to its intensity, though we do not have a variant simulation to show this explicitly. The cosmic average amplitude of the ionizing background radtiation at $z=0$ is not as well constrained as it is at higher redshifts \citep[$2<z<4$, see e.g.][]{2020MNRAS.493.1614F}, and it is likely higher than the average near massive, star-forming galaxies like those in COLIBRE hosting the largest populations of circumgalactic H\,\textsc{i} clouds. This could also lead to producing too many clouds. We conclude that the simulations' prediction of a large population of circumgalactic H\,\textsc{i} clouds likely rests on uncertain assumptions and approximations used in COLIBRE to predict the transition from ionized to neutral gas, and is resolution-dependent. It may be possible to rectify this discrepancy by adjusting these while preserving the overall successes of the model.

\section{Summary and conclusions} \label{sec:conclusions}

The COLIBRE simulations promise abundant avenues for the study of H\,\textsc{i} in statistically representative samples of galaxies. They track gas cooling to low ($\approx 10\,\mathrm{K}$) temperatures and implement an explicit treatment of the formation and destruction of H\,\textsc{i}, including non-equilibrium effects for the most important reactions. The influence of dust on the chemical network is also captured through coupling to a novel dust model tracking its formation, growth and destruction. Unlike many earlier large-volume simulations, COLIBRE resolves a multiphase ISM (but not the internal structure of the cold neutral medium and molecular clouds) and explicitly outputs self-consistent H\,\textsc{i} fractions for each particle, obviating the need for empirically calibrated and/or approximate post-processing methods to identify which gas particles carry H\,\textsc{i}. We have carried out an initial exploration of the H\,\textsc{i} properties of $z\approx0$ galaxies in the COLIBRE simulations. Our main conclusions are:
\begin{itemize}
\item About $10$ resolution elements worth of H\,\textsc{i} ($M_{\mathrm{HI}}\approx 2 \times 10^{8}$, $2 \times 10^7$ and $2 \times 10^{6}\,\mathrm{M}_{\odot}$ at the m7, m6 and m5 resolution levels, respectively) is a good rule of thumb for a lower threshold to select galaxies reasonably well-converged in $M_{\mathrm{HI}}$ across the three resolution levels. However, where stellar properties are involved (e.g. $M_{\mathrm{HI}}/M_{\star}$), using an additional lower threshold of $\approx100$ stellar particles is usually advisable.
\item The simulations reproduce the H\,\textsc{i}-stellar mass relation at $M_{\star}\gtrsim10^{9}\,\mathrm{M}_{\odot}$ across all three resolution levels (Fig.~\ref{fig:Mstar-MHI}). The global number density of H\,\textsc{i} sources is also in good agreement with ALFALFA observations (Fig.~\ref{fig:himf-hiwf}), except at the highest resolution where sources are overproduced by a factor of $\approx 2$ at $w_{50}\lesssim300\,\mathrm{km}\,\mathrm{s}^{-1}$ and $M_{\mathrm{HI}}\geq 3\times10^{9}\,\mathrm{M}_{\odot}$.
\item At the low- and high-mass ends of the distribution the balance between gas-rich and gas-poor galaxies is less well reproduced. There is a deficit of low-H\,\textsc{i} mass ($<3\times10^{8}\,\mathrm{M}_{\odot}$), higher dynamical mass ($w_{50}\gtrsim50\,\mathrm{km}\,\mathrm{s}^{-1}$, not inclination-corrected) galaxies (Fig.~\ref{fig:himf-hiwf-split-m6}, both panels), and a deficit of low stellar mass ($\mathrm{M}_{\star}\lesssim10^{8}\,\mathrm{M}_{\odot}$), H\,\textsc{i}-rich ($M_{\mathrm{HI}}/M_{\star}>1$) galaxies (Figs.~\ref{fig:Mstar-MHI} \& \ref{fig:himf-hiwf-split-mstar-m6}).
\item The simulations reproduce the H\,\textsc{i} gas fractions of massive, quiescent galaxies (Fig.~\ref{fig:massive-quenched}), except for a possible excess of galaxies with $M_{\mathrm{HI}}/M_{\star}>10^{-1}$ (see also Fig.~\ref{fig:himf-hiwf-split-m6}, left panel), but this discrepancy vanishes if galaxies are selected by $\mathrm{NUV}-r$ colour instead of sSFR.
\item The simulations predict a steepening slope of the H\,\textsc{i} mass function below $M_{\mathrm{HI}} \approx 10^{8}\,\mathrm{M}_{\odot}$ (Fig.~\ref{fig:himf-hiwf}). There are tentative hints of this feature in archival data that have so far been overlooked. If borne out, low-mass galaxies (down to $M_{\mathrm{HI}}=10^{6}\,\mathrm{M}_{\odot}$) could make a $10$ times larger contribution to $\Omega_{\mathrm{HI}}$ than implied by the shallower slopes usually inferred, but still a small contribution to $\Omega_{\mathrm{HI}}$ overall (about $2$~per~cent, instead of $0.2$~per cent).
\item COLIBRE galaxies have H\,\textsc{i} discs close to the observed mass-size relation, where size is defined at a density of $1\,\mathrm{M}_{\odot}\,\mathrm{pc}^{-2}$ (Fig.~\ref{fig:MHI-rHI}). They are systematically offset to smaller sizes by $<15$~per~cent and with about twice the observed scatter for the m5 and m6 resolution; at m7 resolution deviations are slightly larger.
\item We compared H\,\textsc{i} surface density profiles of COLIBRE L200m6 galaxies to those measured from FEASTS survey data (Fig.~\ref{fig:sig-profiles}). Many profiles closely resemble those observed. At large radii ($r\gtrsim80\,\mathrm{kpc}$), abundant H\,\textsc{i} clouds throughout the circumgalactic medium (Fig.~\ref{fig:clouds}) around COLIBRE galaxies of $M_{\star}\gtrsim 10^{10}\,\mathrm{M}_{\odot}$ keep their surface densities systematically higher than is observed in the low density ($\Sigma_{\mathrm{HI}}\lesssim10^{-1}\,\mathrm{M}_{\odot}\,\mathrm{pc}^{-2}$) regime. This discrepancy might be possible to remedy through adjustments to the uncertain and approximate treatments of the $z=0$ metagalactic UV/X-ray background and/or of the shielding length for ionizing radiation. The clouds also become less abundant and less massive with increasing numerical resolution.
\item Large ($\gtrsim 1\,\mathrm{kpc}$) central holes in the gas surface density of COLIBRE disc galaxies with $M_{\star}\gtrsim5\times10^{10}\,\mathrm{M}_{\odot}$ are very common (Fig.~\ref{fig:holes}), although their occurence frequency decreases as numerical resolution increases. They are created by AGN feedback and are therefore likely sensitive to its detailed implementation.
\end{itemize}

This work has only scratched the surface of possible studies leveraging the COLIBRE simulations for H\,\textsc{i} science. Given the overall very good agreement with observations of H\,\textsc{i} properties of galaxies at $z\approx 0$ that we have shown, looking into how this has arisen through the evolution of H\,\textsc{i} in galaxies over cosmic history is a natural next step (see also A.~J.~Richings et al, in preparation). The COLIBRE simulations are ideal to help interpret the wealth of 21-cm observational data from campaigns underway (e.g. WALLABY, \citealp{2020Ap&SS.365..118K}; DINGO, \citealp{2020bugm.conf...18R}; LADUMA, \citealp{2016mks..confE...4B}; MIGHTEE-H\,\textsc{i}, \citealp{2021A&A...646A..35M}; CRAFTS, \citealp{2019SCPMA..6259506Z}; FASH\,\textsc{i}, \citealp{2024SCPMA..6719511Z}; FEASTS, \citealp{2023ApJ...944..102W}; and others), and from forthcoming H\,\textsc{i} surveys with the Square Kilometre Array Observatory (SKAO).

\section*{Software}

The following software packages were used in this work:
\textsc{astropy} \citep{2022ApJ...935..167A}, 
\textsc{numpy} \citep{2020Natur.585..357H}, 
\textsc{scipy}\footnote{docs.scipy.org/doc/},
\textsc{swiftsimio} \citep{2020JOSS....5.2430B},
\textsc{swiftgalaxy} \citep{2025JOSS...10.9278O},
\textsc{martini} \citep{2024JOSS....9.6860O},
\textsc{skirt} \citep{2020A&C....3100381C},
\textsc{soap} \citep{2025JOSS...10.8252M},
\textsc{photutils} \citep{2026zndo..19636730B},
\textsc{mpi4py} \citep{2021CSE....23d..47D},
\textsc{unyt} \citep{2018JOSS....3..809G},
\textsc{SoFiA-2} \citep{2015MNRAS.448.1922S,2021MNRAS.506.3962W} \&
\textsc{partridge} (Hu\v{s}ko et al., in preparation).

\section*{Acknowledgements}

Thanks to R.~McGibbon and J.~Helly for scientific computing and software support. Thanks to J.~Wang (\begin{CJK}{UTF8}{gbsn}王菁\end{CJK}) for providing tabulated data from \citet{2025ApJ...980...25W}.
KAO acknowledges support by the Royal Society through a Dorothy Hodgkin Fellowship (DHF/R1/231105) and thanks the researchers at ICRAR, University of Western Australia, where much of this article was drafted, for their hospitality.
FH acknowledges funding from the Netherlands Organization for Scientific Research (NWO) through research programme Athena 184.034.002.
SP acknowledges support by the Austrian Science Fund (FWF) through 10.55776/V982.
RJW acknowledges support from the Forrest Research Foundation through the Forrest Fellowship Program. 
This work used the DiRAC@Durham facility managed by the Institute for Computational Cosmology on behalf of the STFC DiRAC HPC Facility (www.dirac.ac.uk). The equipment was funded by BEIS capital funding via STFC capital grants ST/K00042X/1, ST/P002293/1, ST/R002371/1 and ST/S002502/1, Durham University and STFC operations grant ST/R000832/1. DiRAC is part of the National e-Infrastructure.
This research has made use of the Astrophysics Data System, funded by NASA under Cooperative Agreement 80NSSC21M00561.

For the purpose of open access, the author has applied a Creative Commons Attribution (CC BY) licence to any Author Accepted Manuscript version arising from this submission.

\section*{Data Availability}

The COLIBRE simulation data underpinning this work can be requested for specific projects, subject to approval by the COLIBRE Board. Machine-readable versions of data points and curves appearing in figures in this work are available from KAO (kyle.a.oman@durham.ac.uk) on request. The base \textsc{swift} simulation code \citep{2024MNRAS.530.2378S} is publicly available (\url{www.swiftsim.com}). The COLIBRE-specific modules implemented in \textsc{swift} will be released at a later date alongside a public simulation data release.

\bibliographystyle{mnras}
\bibliography{references}

\appendix

\section{Controlled variations of the COLIBRE model}
\label{app:vars}

Although COLIBRE makes an unambiguous and self-consistent prediction for the H\,\textsc{i} content of each simulation particle, the model still contains uncertain ingredients and parameters. In this Appendix we examine the sensitivity to some of our main findings to model variations where some of the model elements most likely to influence H\,\textsc{i} are varied one at a time.

We make use of five re-simulations that are variations of a fiducial L025m6 simulation run with the same model as the L200m6 simulation. We also compare the two fiducial simulations to monitor the influence of the box size and the sampling noise due to the small number of galaxies (particularly massive ones) in the smaller box. The variant simulations are:
\begin{itemize}
\item L025m6\_noISRF \& L025m6\_ISRFx10, variants where the strength of the interstellar radiation field is set to zero and increased by a factor of $10$, respectively;
\item L025m6\_NEQ, a variant where the full HYBRID-CHIMES network ($157$ species) is treated in non-equilibrium mode;
\item L025m6\_Nshx0p5 \& L025m6\_Nshx2p0, variants where the shielding lengths estimated from the Jeans' length are halved or doubled, respectively. 
\end{itemize}
These and other model variations will be discussed in greater detail in an article by E.~Chaikin et al. (in preparation), who will also show that the variations discussed here do not compromise the good agreement with the calibration targets (galaxy stellar mass function and stellar mass-size relation).

\begin{figure}
\includegraphics[width=\columnwidth]{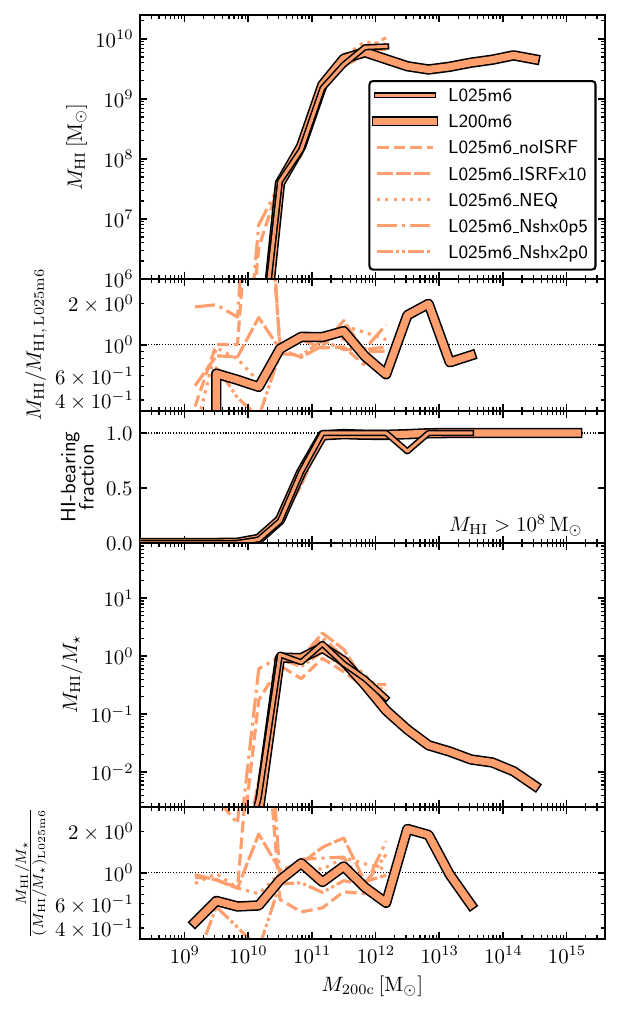}
\caption{Similar to Fig.~\ref{fig:Mhalo-MHI}. The heavy curves showing results from the L200m6 simulation are repeated from that figure. The thinner solid lines show the results from the smaller L025m6 volume run with the same COLIBRE model at the same resolution. Broken lines show variations increasing the strength of or switching off the interstellar radiation field, enabling the full HYBRID-CHIMES non-equilibrium chemistry network, and increasing/decreasing the Jeans length-based shielding column density estimation, as labelled (see Appendix~\ref{app:vars}). In some cases the broken lines are so similar to the solid lines that they are hidden behind them. The median H\,\textsc{i} mass at fixed halo mass (upper panel) is remarkably similar across all of these model variations. The second panel shows the ratio of this median H\,\textsc{i} mass for all other models with resect to that for the L025m6 simulation. Comparing to the thick line in this panel shows that differences are consistent with sampling noise due to the smaller number of galaxies in the smaller volume. The fraction of H\,\textsc{i}-bearing galaxies ($M_{\mathrm{HI}}\geq10^{8}\,\mathrm{M}_{\odot}$) as a function of halo mass (third panel) is essentially identical across all model variants. The H\,\textsc{i}-to-stellar mass ratio (fourth panel, with the ratio with respect to the L200m6 simulation in the lower panel) is sensitive to these model changes, most strongly to modifying the strength of the interstellar radiation field (L025m6\_noISRF and L025m6\_ISRFx10, separated by up to a factor of $\approx 4$ near $M_{\mathrm{200c}}=10^{11}\,\mathrm{M}_{\odot}$). These differences are driven by changes in the stellar-to-halo mass relation, not the H\,\textsc{i}-to-halo mass relation, which the upper panels show is unchanged.}
\label{fig:Mhalo-MHI-mv}
\end{figure}

Fig.~\ref{fig:Mhalo-MHI-mv} is similar to Fig.~\ref{fig:Mhalo-MHI} that showed the dependence of H\,\textsc{i} mass on halo mass, but comparing the different model variants. Two panels showing the ratio of the median H\,\textsc{i} mass and H\,\textsc{i}-to-stellar mass ratio of each simulation variant with respect to those in the L025m6 simulation have been added to more clearly show the differences. The median H\,\textsc{i} mass as a function of halo mass (upper two panels) is remarkably insensitive changing the ISRF, to switching on the full non-equilibrium chemistry network, and to changing the shielding length across the range $10^{10.5}\lesssim M_{\mathrm{200c}}/\mathrm{M}_{\odot}\lesssim10^{11.5}$. At lower halo masses the median plummets, and this is reproduced in all models but sometimes shifted by one bin -- we do not regard this as a significant change. The $(25\,\mathrm{cMpc})^{3}$ volume does not contain enough galaxies to compare more massive haloes. The differences at $M_{\mathrm{200c}}\gtrsim10^{11.5}\,\mathrm{M}_{\odot}$ for simulations where the physics are varied are comparable to those between the L200m6 and L025m6 simulation, so we attribute them to sampling noise. The fraction of haloes containing at least $10^{8}\,\mathrm{M}_{\odot}$ of H\,\textsc{i} as a function of halo mass (third panel) is essentially identical across all simulation variations. On the other hand, the H\,\textsc{i} gas fraction ($M_{\mathrm{HI}}/M_{\star}$) as a function of halo mass does show some systematic differences as the physics are varied: increasing the ISRF strength or decreasing the shielding length increases the H\,\textsc{i} gas fractions, and switching off the ISRF or increasing the shielding length decreases them. At first this may seem counter-intuitive, but the upper panels make it clear that the H\,\textsc{i} content of haloes at $z=0$ is essentially unchanged -- it is the stellar masses that are different. A stronger ISRF or decreased shielding length would intuitively suppress star formation, which explains the trends.

\begin{figure}
\includegraphics[width=\columnwidth]{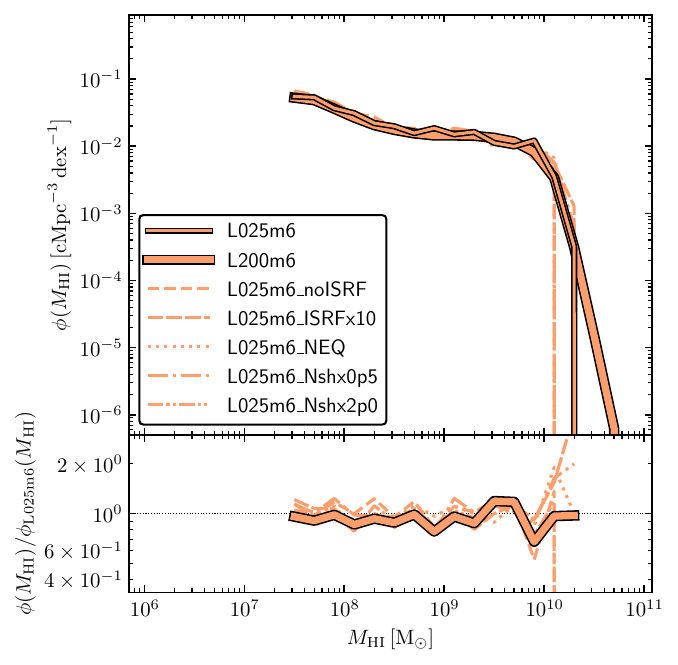}
\caption{Similar to Fig.~\ref{fig:himf-hiwf}, but only showing the H\,\textsc{i} mass function. The heavy line showing the H\,\textsc{i} mass function in the L200m6 simulation is similar to that in Fig.~\ref{fig:himf-hiwf}, but uses all galaxies in the simulation instead of only those in a mock survey volume. The thinner solid line shows the H\,\textsc{i} mass function from the smaller L025m6 volume run with the same COLIBRE model at the same resolution. Broken lines show variations increasing the strength of or switching off the interstellar radiation field, enabling the full HYBRID-CHIMES non-equilibrium chemistry network, and increasing/decreasing the Jeans length-based shielding column density estimation, as labelled. The broken lines are mostly hidden behind the heavy line. The ratio with respect to the L025m6 H\,\textsc{i} mass function is plotted for all other models in the lower panel. None of the variations produce differences of more than about $20$~per~cent compared to the fiducial model in the L025m6 volume, except at the high-mass end where the smaller box used in the model variants causes a truncation. The thick line shows that these differences are consistent with sampling noise due to the smaller number of galaxies in the smaller volume.}
\label{fig:himf-mv}
\end{figure}

Fig.~\ref{fig:himf-mv} shows the H\,\textsc{i} mass function, similar to Fig.~\ref{fig:himf-hiwf}, but showing only the mass function and only the volume-limited (and not the mock-observed flux-limited) sample. A panel has been added to show fractional changes with respect to the L025m6 simulation more clearly. None of the variant simulations considered differs from the fiducial L025m6 simulation by more than the difference between the L200m6 and L025m6 simulations, so we attribute these to sampling noise. The truncation of the mass function around $2\times10^{10}\,\mathrm{M}_{\odot}$ for all of the simulations in the $(25\,\mathrm{cMpc})^{3}$ volume arises because the volume is too small to sample the rarest, most H\,\textsc{i}-rich galaxies.

\begin{figure}
\includegraphics[width=\columnwidth]{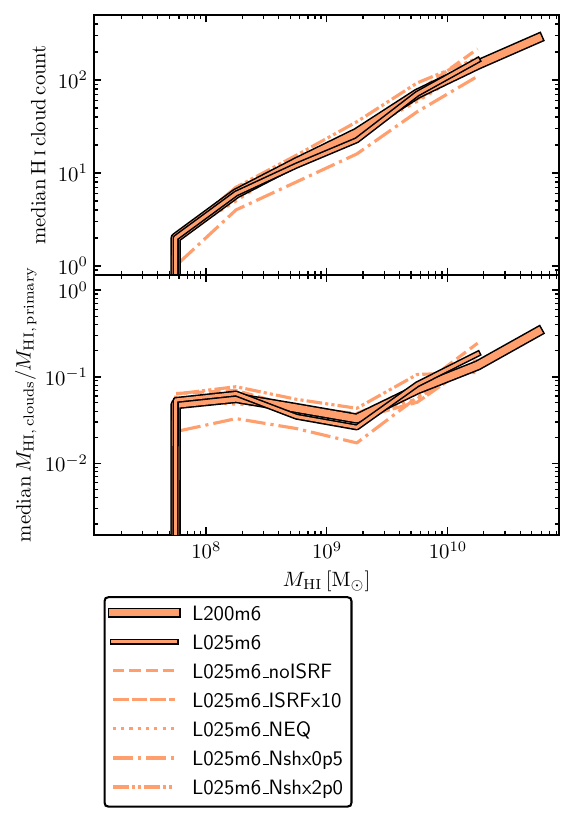}
\caption{Similar to Fig.~\ref{fig:clouds}. The heavy lines showing results from the L200m6 simulation are repeated from that figure. The thinner solid lines show the results from the smaller L025m6 volume run with the same COLIBRE model at the same resolution. Broken lines show variations increasing the strength of or switching off the interstellar radiation field, enabling the full HYBRID-CHIMES non-equilibrium chemistry network, and increasing/decreasing the Jeans length-based shielding column density estimation, as labelled. Several of the broken lines are so similar to the L200m6 simulation that they are hidden behind its line, only the shielding length variations (L025m6\_Nsh0p5 and L025m6\_Nsh2p0) impact the abundance and H\,\textsc{i} mass in circumgalactic clouds, by about a factor of $2$ between halving and doubling the shielding length (factor of $4$ change).}
\label{fig:clouds-mv}
\end{figure}

Finally, Fig.~\ref{fig:clouds-mv} shows the same assessment of the abundance and mass contained in circumgalactic H\,\textsc{i} clouds as in Fig.~\ref{fig:clouds}, but comparing the different model variants. Changing the ISRF or switching on full non-equilibrium chemistry does not affect the abundance or mass contained in the circumgalactic clouds (at least, no more than the difference between the L200m6 and L025m6 simulations, that we use as a threshold to assess how large differences must be to be considered significant). However, increasing the shielding length by a factor of $2$ (L025m6\_Nshx2p0) slightly increases the number of circumgalactic clouds and the total mass contained in them (except at $M_{\mathrm{HI}}\gtrsim10^{10}\,\mathrm{M}_{\odot}$ where the mass not measurably different), while reducing the shielding length by half (L025m6\_Nshx0p5) decreases their abundance and H\,\textsc{i} mass content by a factor of nearby $2$ (again except at $M_{\mathrm{HI}}\gtrsim10^{10}\,\mathrm{M}_{\odot}$). Taken all together, this suggests that the size of the clouds is limited not by the ISRF but likely rather the UV background radiation field. We have not explored model variations that increase the intensity of metagalactic UV/X-ray background (a factor of $2$ increase at $z=0$ is plausible, especially near massive galaxies that provide a local source of ionizing radiation), or even smaller smoothing lengths that are physically appropriate for pressure-confined clouds. Since clouds like those around galaxies in COLIBRE do not seem to be observed (Sec.~\ref{sec:cgm-clouds}, see also Sec.~\ref{sec:SigmaHI}), at least not in the same abundance, understanding their detailed origin in the simulations will be an interesting topic for further study and may suggest ways to refine future simulation models.

\section{MHONGOOSE-like mock observations} \label{app:mocks-mhongoose}

The visualisations in Figs.~\ref{fig:visual-impression-discs} \& \ref{fig:visual-impression-groups} include both H\,\textsc{i} moment maps and optical images. The procedures used to generate these are described in Secs.~\ref{app:mocks-meerkat} \& \ref{app:mocks-hst}, respectively. Both procedures have a common beginning: we use the \textsc{swiftgalaxy} utility \citep{2025JOSS...10.9278O} to load all particles within a sphere of $500\,\mathrm{kpc}$ radius around the chosen centre point (usually the most bound particle in the target galaxy, except for the group in field~D where a target location was chosen by hand). The reference frame is shifted in velocity to match the mass-weighted mean velocity of particles bound to the target galaxy. The particles are then rotated to achieve the desired viewing angle before being passed to one of the processes described in the next subsections.

\subsection{\texorpdfstring{H\,\textsc{i}}{HI} moment maps} \label{app:mocks-meerkat}

The \textsc{martini} code\footnote{Version 2.1.15, git commit ID bedf5a9.} \citep{2019MNRAS.482..821O,2024JOSS....9.6860O} was configured with the source placed at the distance listed in Table~\ref{tab:img-props} with $2048^2$, $2\,\mathrm{arcsec}$ pixels, $1024$, $1\,\mathrm{km}\,\mathrm{s}^{-1}$ channels (except field C where half as many, twice as wide channels were used to reduce memory usage), and the sight line centred on the target. Particles are injected into the datacube grid using an SPH `scatter' operation \citep{2012JCoPh.231..759P} onto the spatial grid, assuming the quartic spline SPH kernel used in the COLIBRE simulations and the smoothing lengths of the gas particles recorded in the snapshots, and a Gaussian line profile discretized onto the velocity grid -- the line width is tied to the particle temperatures using \textsc{martini}'s `thermal' broadening option. Gaussian white noise is then added to the cube so that it will have an amplitude of $0.2\,\mathrm{mJy}\,\mathrm{beam}^{-1}$ after convolution with the beam, then the cube is convolved with an $8\,\mathrm{arcsec}$ (FWHM) circular beam. The initial data cube is padded so that the convolution yields correct values within the region of interest; the pad is then discarded. The mock observation is written out in FITS format.

The FITS cube is then passed to the \textsc{SoFiA-2} source finding code\footnote{Version 2.6.0.} with its default configuration and output of moment maps and source cubelets switched on. These moment maps are the ones plotted in Figs.~\ref{fig:visual-impression-discs} \& \ref{fig:visual-impression-groups} (the cubelets, i.e. omitting emission from other sources, are used in the lower panels of each set).

\subsection{Optical images} \label{app:mocks-hst}

The \textsc{partridge} code\footnote{No semantic versioning was established for \textsc{partridge} when the analysis was carried out. The version used is a customised variant (git commit ID: 5453f5a) of a version under development that supports interfacing with \textsc{swiftgalaxy}.} (Hu\v{s}ko et al., in preparation) was used to produce the optical images. It was configured to have the same field of view used for the \textsc{martini} mock observation of the same object and $512^2$ pixels. \textsc{partridge} works by rendering the stellar and dust properties onto a 3D spatial grid, then applying a 1-dimensional radiative transfer method (i.e. dust attenuation) while collapsing the grid into a 2D image. \textsc{Partridge} works with cubic volumes, so we image multiple cubic volumes along the line of sight to match the depth covered in the H\,\textsc{i} mock observations, then stack the images. There are no massive (e.g. enough to cause significant dust attenuation) galaxies in separate cubic volumes but overlapping along the line of sight in these images. We set the surface brightness threshold to $1\,\mathrm{L}_{\odot}\,\mathrm{pc}^{-2}$ and supersampled the final (full colour) images onto a 4-times finer pixel grid using spline interpolation.

\section{ALFALFA-like mock observations} \label{app:mocks-alfalfa}

In Sec.~\ref{sec:himf-hiwf} we discussed the COLIBRE H\,\textsc{i} mass functions and velocity width functions derived from an ALFALFA-like mock survey -- here we give further details of the mock survey procedure. We place an `observer' at the origin of the simulation cube coordinate frame (this is an arbitrary choice) and consider a sight line along the $x$-axis of the cube. A cone with a $200\,\mathrm{Mpc}$ length, a spherical cap with the same radius, and an opening angle of $\approx 46.8\,\mathrm{deg}$ \citep[the exact survey footprint tabulated in tables~D1 \& D2 of][is used to evaluate the area and equivalent opening angle]{2018MNRAS.477....2J} is placed with its tip on the observer. The periodic simulation volume is then wrapped and/or tiled as necessary to cover the conical volume (this tiling has no serious consequences for our results as discussed in Sec.~\ref{sec:himf-hiwf}). The positions of galaxies in the volume are mapped to distances and sky coordinates in the observer's frame. The \textsc{martini} code\footnote{Version 2.1.11, git commit ID 5ffe2b7.} \citep{2019MNRAS.482..821O,2024JOSS....9.6860O} was used in its `GlobalProfile' mode to produce a mock H\,\textsc{i} spectrum of each galaxy. In this mode the on-sky positions of particles are not used, but the sight line to each particle is still used to determine its Doppler shift and discretize its spectrum onto the velocity grid with a Gaussian line profile whose width is tied to the particle temperatures using \textsc{martini}'s `thermal' broadening option. We do not add noise since the data are intended for comparison of $w_{50}$ line widths with ALFALFA `Code 1' sources that have signal-to-noise ratios $>6.5$.

We include all particles flagged as gravitationally bound by the halo finder in the spectra. This implies an unrealistically perfect ability to separate emission from galaxies closely spaced in position and velocity, especially when considering the $2\,\mathrm{arcmin}$ Arecibo telescope beam. We approximate the effect of `source confusion' by searching for sources in our mock catalogues (before applying the selection function) that are separated by less than the sum of their $w_{50}$ line widths and less than $2\,\mathrm{arcmin}$. Such sources have their H\,\textsc{i} and stellar masses summed, their spectra summed and $w_{50}$ re-measured, then a new source with these new properties is inserted into the catalogue and its constituent sources deleted. This search, merge and replacement is repeated recursively until convergence. We find that the influence of confusion is small, especially in terms of counts in the catalogue of mock detections, and does not impact the interpretation presented in Sec.~\ref{sec:himf-hiwf}.

\section{Additional decompositions of the \texorpdfstring{H\,\textsc{i}}{HI} mass and velocity width functions} \label{app:himf-hiwf}

\begin{figure*}
  \includegraphics[width=\textwidth]{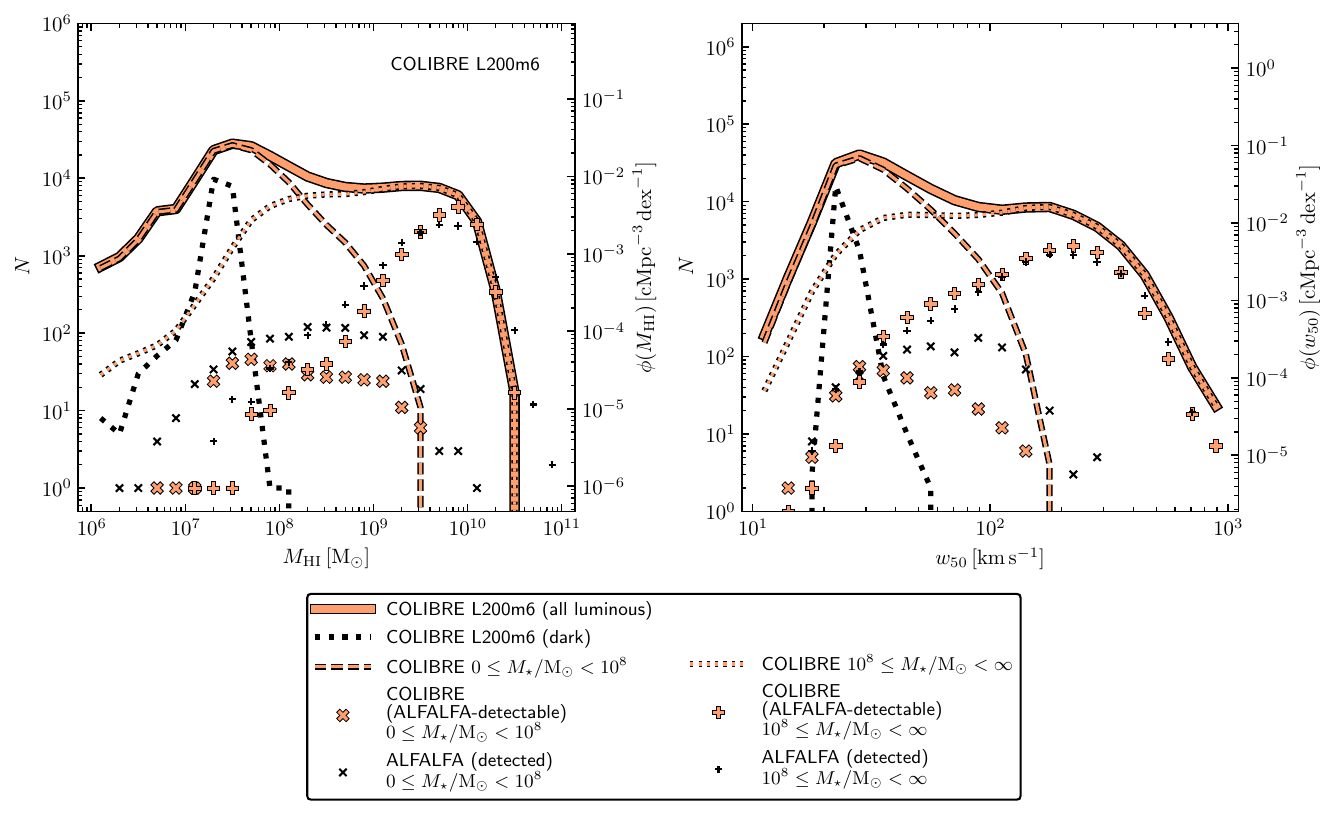}
  \caption{Similar to Fig.~\ref{fig:himf-hiwf-split-m6}, but for the L200m6 simulation and separating the contributions of galaxies above and below $M_{\star}=10^8\,\mathrm{M}_{\odot}$. Galaxies with no stars (`dark') are shown separately with the dotted black line.}
  \label{fig:himf-hiwf-split-mstar-m6}
\end{figure*}

In Fig.~\ref{fig:himf-hiwf-split-m6} we showed the decomposition of the L200m6 simulation H\,\textsc{i} mass function into low- and high-$w_{50}$ components, and the H\,\textsc{i} velocity width function into low- and high-$M_{\mathrm{HI}}$ components. In Fig.~\ref{fig:himf-hiwf-split-mstar-m6}, we include alternative decompositions into low- ($\leq 10^{8}$) and high-stellar mass ($>10^{8}\,\mathrm{M}_{\odot}$) components. We use a reduced mock survey footprint of $4051\,\mathrm{deg}^{2}$ to enable direct comparison with the ALFALFA sample with cross-matched SDSS stellar mass measurements of \citet[][stellar masses are again corrected according to their eq.~(3)]{2020AJ....160..271D}. The H\,\textsc{i} mass and velocity width functions of simulated galaxies with no optical counterparts in the same mock survey volume are shown with dotted black lines. The nature of these galaxies will be explored in detail by F.~Turini et al. (in preparation). The other curves do not include the contribution from these galaxies. We draw similar conclusions as from Fig.~\ref{fig:himf-hiwf-split-m6}. The higher stellar mass subsample has a deficit of galaxies compared to those observed at $M_{\mathrm{HI}}\lesssim10^{9}\,\mathrm{M}_{\odot}$ and an excess near the `knee' at $M_{\mathrm{HI}}\approx10^{10}\,\mathrm{M}_{\odot}$ while the lower stellar mass subsample \citep[noting that a $M_\star=10^{8}\,\mathrm{M}_{\odot}$ galaxy is only resolved with about $50$ stellar particles at this resolution level and that the abundance of these galaxies is not converged when the resolution is increased;][]{2026MNRAS.548ag375S} has a deficit relative to observations at $M_{\mathrm{HI}}\gtrsim10^{7.5}\,\mathrm{M}_{\odot}$ and $w_{50}\gtrsim 30\,\mathrm{km}\,\mathrm{s}^{-1}$.

\section{FEASTS-like mock observations}\label{app:mocks-feasts}

In Sec.~\ref{sec:SigmaHI} we compared the surface density profiles of COLIBRE galaxies with those observed for the FEASTS survey using the FAST telescope. Here we describe the mock-observing procedure. We take a matched sample approach. For each observed galaxy we select a simulated galaxy with a very similar H\,\textsc{i} mass and stellar mass by finding the galaxy that minimizes the distrance metric $\sqrt{\left(\Delta \log_{10}(M_{\star})\right)^2 + \left(\Delta \log_{10}(M_{\mathrm{HI}})\right)^2}$, where $\Delta \log_{10}M_{\star}$ and $\Delta \log_{10}M_{\mathrm{HI}}$ are the differences on a logarithmic scale with respect to the observed stellar and H\,\textsc{i} masses, respectively. For the observed galaxies, we use stellar mass estimates from \citet{2019ApJS..244...24L}, except for NGC 4496(A) which is absent from that catalogue; we adopt the value from \citet{2022MNRAS.512.3284S} for it. H\,\textsc{i} masses are provided by \citet{2025ApJ...980...25W}. The L200m6 simulation contains enough galaxies that we find an essentially exact match for each observed galaxy (maximum offset: $0.03\,\mathrm{dex}$). One match is a duplicate; we replace one instance with the second-nearest neighbour. We have visually inspected $(300\,\mathrm{kpc})^2$ images of the H\,\textsc{i} density distribution around the simulated galaxies along $3$ orthogonal projection axes, one of which is parallel to the stellar angular momentum axis calculated from stars within a $10\,\mathrm{kpc}$ spherical aperture. $9$ galaxies in our matched sample show clear signs in their H\,\textsc{i} morphology of interaction with companions; \citet{2025ApJ...980...25W} identify $11$ interacting galaxies in their sample.

We next place each simulated galaxy at the same distance as its observed counterpart from a notional observer and orient it so that the angular momentum vector of stars in the central $10\,\mathrm{kpc}$ is parallel to the line of sight (i.e. face-on). We then incline each galaxy by the inclination angle reported by \citet{2025ApJ...980...25W}. We comment further on inclination effects below. We next use the \textsc{martini} code\footnote{Version 3.0.0, git commit ID 9e4f320.} \citep{2019MNRAS.482..821O,2024JOSS....9.6860O} to create mock data cubes. We match the characteristics of the FEASTS observations: pixels are $30\,\mathrm{arcsec}$ on a side, channels are $1.61\,\mathrm{km}\,\mathrm{s}^{-1}$ wide, the mock cubes are convolved with a $3.24\,\mathrm{arcsec}$ Gaussian beam, and noise is added (before the beam convolution step) to obtain a RMS noise level of $1\,\mathrm{mJy}\,\mathrm{beam}^{-1}$ in the final cubes. The size of the cubes (in terms of pixel count) is adapted to the distance of the sources so that they are $\geq300\,\mathrm{kpc}$ on a side. 

We run the \textsc{SoFiA-2} source finder with the same configuration used for FEASTS data \citep[see][]{2025ApJ...980...25W} and select the source with the highest H\,\textsc{i} flux for further analysis. We use the `aperture\_photometry' utility from the \textsc{photutils} package\footnote{Version 3.0.0.} \citep{2026zndo..19636730B} to extract surface density profiles from the zeroth moment maps produced by \textsc{SoFiA-2} in elliptical ring apertures of $1\,\mathrm{kpc}$ width centred on the most bound particle in the target galaxy. The ellipses have aspect ratios set by the inclination of the galaxy, and the surface density profile is corrected for inclination assuming a thin disc geometry (multiplication by the cosine of the inclination), as is done for the FEASTS surface density profiles.

We note that our treatments of centres and inclinations are slightly inconsistent with observations. A more accurate matched sample would centre on an observable feature and have the inclination adjusted until the apparent inclination \citep[as measured from the aspect ratio of the $N_{\mathrm{HI}}=10^{19}\,\mathrm{cm}^{-2}$ isophote as in][]{2025ApJ...980...25W} matched that observed. This proved too unwieldly to implement, so we instead incline by the observed inclination with respect to the stellar angular momentum axis as described above. We have also extracted H\,\textsc{i} surface density profiles from the same mock moment maps assuming an inclination of $0$ for all galaxies. This increases the central surface brightnesses somewhat (by a typical factor of $\approx(\cos 60^{\circ})^{-1}$, or $\approx 0.3\,\mathrm{dex}$), as expected, but even this radical change would not qualitatively alter the main conclusions pertaining to the outer profiles reached in Sec.~\ref{sec:SigmaHI}.

\bsp
\label{lastpage}
\end{document}